\def\gsim{~\rlap{$>$}{\lower 1.0ex\hbox{$\sim$}}}
\def\lsim{~\rlap{$<$}{\lower 1.0ex\hbox{$\sim$}}}

\def\h2o{\rm{H_{2}O}}
\def\mh2{\rm{H_{2}}}

\def\co2{\rm{CO_{2}}}
\def\ch4{\rm{CH_{4}}}
\documentclass[12pt, preprint]{aastex}
\usepackage{graphicx}
\usepackage{natbib,color,lscape}
\usepackage{subfigure}
\usepackage{threeparttable}
\usepackage{url}
\usepackage{subfigure}
\usepackage{threeparttable}
\usepackage{rotating}

\begin{document}
\title{Habitable Zones Around Main-Sequence Stars: New Estimates}
\author{Ravi kumar Kopparapu\altaffilmark{1,2,3,4}, 
        Ramses Ramirez\altaffilmark{1,2,3,4}, 
        James F. Kasting\altaffilmark{1,2,3,4}, 
        Vincent Eymet\altaffilmark{5}, 
        Tyler D. Robinson\altaffilmark{2,6,7}
        Suvrath Mahadevan\altaffilmark{4,8}
        Ryan C. Terrien\altaffilmark{4,8}
        Shawn Domagal-Goldman\altaffilmark{2,9}
        Victoria Meadows\altaffilmark{2,6}
        Rohit Deshpande\altaffilmark{4,8}}
\altaffiltext{1}{Department of Geosciences, Penn State University, 443 
Deike Building, University Park, PA 16802, USA}
\altaffiltext{2}{NASA Astrobiology Institute's Virtual Planetary Laboratory}
\altaffiltext{3}{Penn State Astrobiology Research Center, 2217 Earth and Engineering Sciences Building
University Park, PA 16802}
\altaffiltext{4}{Center for Exoplanets \& Habitable Worlds, The Pennsylvania State University, University
Park, PA 16802}
\altaffiltext{5}{Laboratoire d'Astrophysique de Bordeaux, Universite de Bordeaux 1, UMR 5804}
\altaffiltext{6}{Astronomy Department, University of Washington, Box 351580, Seattle, WA 98195-1580, USA}
\altaffiltext{7}{University of Washington Astrobiology Program}
\altaffiltext{8}{Department of Astronomy \& Astrophysics, The Pennsylvania State University, 525 Davey
Laboratory, University Park, 16802, USA}
\altaffiltext{9}{Planetary Environments Laboratory, NASA Goddard Space Flight Center}

\begin{abstract}
Identifying terrestrial planets in the habitable zones (HZs) of
 other stars is one of the primary goals of
ongoing radial velocity and transit exoplanet surveys and proposed future space missions. 
Most current estimates of the boundaries of the HZ are based on  1-D, cloud-free, 
climate model 
calculations 
by \cite{Kasting1993}. However, this model used band models which were based on older
HITRAN and HITEMP line-by-line databases.
The inner edge of the HZ in \cite{Kasting1993} model was 
determined by loss of water, and 
the outer edge was determined by the maximum greenhouse provided by a $\co2$ atmosphere.
A conservative
estimate for the width of the HZ from this model in our Solar system is 0.95-1.67 AU.

Here, an updated 1-D
radiative-convective, cloud-free climate model is used to obtain new estimates for HZ widths 
around F, G, K and M stars. New $\h2o$ and $\co2$ absorption coefficients, 
derived from the HITRAN 2008 and HITEMP 2010
line-by-line databases, are important improvements to the climate model.
According to the new model, the water loss (inner HZ) and maximum greenhouse (outer HZ) limits 
for our Solar System are at
$0.99 $ AU and $1.70$ AU, respectively, suggesting that the present Earth lies near the inner edge. Additional calculations
are performed for stars with effective temperatures 
between $2600$ K and $7200$ K, and the results are presented in parametric form, making them easy to apply to actual stars. 
The new model indicates that, near the inner edge of the HZ,
 there is no clear distinction between runaway greenhouse and water loss limits
for stars with T$_{eff} \lesssim 5000$ K which has implications for ongoing planet searches
around K and M stars.
To assess the potential habitability of extrasolar terrestrial planets, we propose using stellar flux
incident on a planet rather than equilibrium temperature. This removes the dependence on planetary (Bond) albedo, 
which varies depending upon the host star's spectral type. We suggest that conservative estimates of the HZ 
(water loss  and maximum greenhouse limits) should be
used for current RV surveys and {\it Kepler} mission to obtain a lower limit on $\eta_{\oplus}$, so that 
future flagship missions like {\it TPF-C} and {\it Darwin} are not undersized.
Our model does not include the radiative effects of clouds; thus,
the actual HZ boundaries may extend further in both directions than the estimates just given. 
\end{abstract}
\keywords{stars: planetary systems}

\maketitle

\section{Introduction}
\label{intro}

As of November, 2012, more than 800 extra-solar planetary systems have been
detected\footnote{\url {exoplanets.org}}, and  $> 2000$ additional candidate systems
from the {\it Kepler} mission are waiting to be confirmed \citep{Batalha2012}. 
One of the primary goals of the ongoing radial velocity (RV) and transit surveys is
to identify a terrestrial mass planet ($0.3 - 10$M$_{\oplus}$) in the so-called 
Habitable Zone (HZ), which
is traditionally defined as the circumstellar region in which a terrestrial-mass 
planet with a $\co2$-$\h2o$-N$_{2}$ atmosphere can sustain liquid water on its surface\footnote{
\cite{Abe2011} studied habitability of  water-limited 'land' planets and found that they could
 remain habitable much closer to their host stars. However, \cite{Abbot2012} found that
a waterworld would have a narrower HZ owing to lack of weathering-climate feedback.}
\citep{Huang1959, Hart1978, Kasting1993, Underwood2003, Selsis2007b, Kaltenegger2011b,PG2011}.
Several potential HZ planet candidates have already been detected, 
\citep{Udry2007, Pepe2011a, Borucki2011, Bonfils2011, Borucki2012, Vogt2012, Tuomi2012a, Tuomi2012b}
 and it is
expected that this number will greatly increase as time passes \citep{Batalha2012}. In the near 
future we may be able to study  habitable planets
orbiting nearby M stars. These planets are relatively close to their parent stars,
leading to shorter orbital periods and an increase in the probability of a transit.
NASA's {\it James Webb Space Telescope} (JWST), 
scheduled to launch in 2018, is considered to be marginally capable of obtaining a transit 
spectrum of an Earth-like planet orbiting a late M dwarf \citep{Clampin2007, 
KT2009,Deming2009}. 
 Several other surveys 
are either underway \citep[MEARTH]{Nutzman2008} or getting ready to be commissioned
\citep[HPF]{Suvrath2012} in an attempt to discover rocky planets in the HZs of
low mass stars.

The HZ limits that were cited in many recent  discoveries were obtained from 
 1-D radiative-convective, cloud-free climate model 
calculations by \cite{Kasting1993}. For our Sun, these authors estimated the boundaries 
of the HZ to be
$0.95$ AU for the inner edge and $1.67$ AU for the outer edge. These values represent the 
``water loss'' and ``maximum greenhouse'' limits, respectively. Other, less conservative
limits for the inner edge are the ``runaway greenhouse'' and ``recent Venus'' limits. The
latter estimate is empirical, based on the inference that Venus has not had liquid water 
on its surface for at least the last 1 billion years \citep{SH1991}. For the
outer edge, there is a corresponding ``early Mars'' empirical estimate, based on the 
inference that Mars did have liquid water on its surface 3.8 billion years ago. (The
``1st $\co2$ condensation'' limit of \cite{Kasting1993}, should now be disregarded, as it
has been shown that $\co2$ clouds generally warm a planet's climate \citep{FP1997}).
Some studies have investigated the effects of clouds on planetary emission spectra of Earth-like
planets in a 1D model \citep{Kitzmann2011a,Kitzmann2011b}, while others studied the habitability of
specific systems, particularly Gl 581, in 1D \citep{Wordsworth2010, vparis2011a,Kaltenegger2011a}
and 3D \citep{Wordsworth2011, RayP2011}.
 Several other studies \citep{Underwood2003, Selsis2007b} parameterized these results to estimate relationships between HZ boundaries and stellar parameters for stars of different spectral types.

Although these studies provided useful estimates of the HZ width, the \cite{Kasting1993}
model has become outdated, for several reasons:
\begin{enumerate}
\item \cite{Kasting1993} used `band models'\footnote{See Appendix B of \cite{Kasting1988} for a 
detailed description of the band model.} for $\h2o$ and $\co2$ absorption in the thermal-infrared.
 These coefficients were considered valid up to $\sim 700$ K. These coefficients were later 
replaced \citep{Mischna2000} by coefficients generated using the 
correlated-{\it k} technique \citep{Mlawer1997, Kato1999}. A
line-by-line (LBL) radiative transfer model, in this case LBLRTM \citep{CI1995},
was used to generate detailed spectra for $\h2o$ and $\co2$ at a variety of different
temperatures and pressures. Once the detailed spectra were calculated, separate broad-band
k-coefficients for both $\h2o$ and $\co2$ were generated by R. Freedman using standard procedures.
But these coefficients were only derived for temperatures $< 350$ K and should therefore underestimate 
thermal-IR absorption in warm, moist greenhouse atmospheres. (This prediction was verified by 
direct experimentation with that model.) Furthermore, the coefficients adopted by \cite{Mischna2000}
 and used
in subsequent climate modeling studies by the Kasting research group
were obtained using HITRAN 1996 database and had not been updated since then. 

\item Recent studies \citep{Halevy2009, Wordsworth2010} have pointed that the
 \cite{Kasting1993} model may have significantly 
overestimated absorption of
 thermal-IR radiation by collision-induced absorption (CIA) bands of $\co2$, which may affect
the outer edge of the HZ. 

\item The \cite{Kasting1993} calculations spanned stellar effective
temperatures from $7200$ K to $3700$ K, corresponding approximately to stellar classes F0
to M0. Stellar effective temperature affects the HZ boundaries because the radiation from
F stars is bluer relative to that from the Sun, whereas the radiation from K and M
stars is redder, and this affects calculated planetary albedos.
The HZ limits from \cite{Kasting1993} model do not include  M stars with 
effective temperatures lower than $3700$ K. As pointed out above, such stars  
are promising candidates for current observational surveys because 
their HZs are closer to the star. Therefore, potential rocky planets in the HZs will have shorter 
orbital periods and higher probability of transit.
\end{enumerate}

In this paper we address all the above major issues with the goal of deriving new, improved estimates
for the boundaries of the HZ.
 The outline of the paper is as follows: In \S\ref{model} we describe
our 1-D cloud-free climate model, corresponding model updates and model validation with 
other studies.
In \S\ref{results} we present results from our climate model and discuss various HZ limits
for our Earth. \S\ref{FGKM} presents HZ boundaries around F, G, K and M spectral stellar 
spectral types, then provides a generalized expression to calculate HZ boundaries and compares
these boundaries with previous studies. We discuss the implications of these new results 
for currently known exoplanet planetary systems in \S\ref{discussion} 
and present our conclusions in \S\ref{conclusions}.

\section{Model description}
\label{model}
We used a one-dimensional, radiative-convective, cloud-free climate model based on 
\cite{Kasting1988} for the inner edge of the HZ (IHZ) and \cite{Kasting1991} for the 
outer edge of the HZ (OHZ) calculations. 
Following \cite{Kasting1993}, we assumed an Earth-mass planet with an $\h2o$ (IHZ) or $\co2$
(OHZ) dominated atmosphere for our base model. Sensitivity studies for different planetary masses are described in the 
following section. Both the inner and outer edge calculations relied on so-called ``inverse
climate modeling'', in which the surface temperature is specified, and the model is used to
calculate the corresponding solar flux needed to sustain it. To do this, the atmosphere was divided into $101$ layers, and a 
specific pressure-temperature profile was assumed. For the inner edge, this consisted of a moist pseudoadiabat
extending from the surface up to an isothermal (200 K) stratosphere. Methodology for calculating the 
pseudoadiabat was taken from Appendix A of \cite{Kasting1988}. The surface temperature was varied 
from 200-2200 K
during the course of the calculations. For the outer edge, the surface temperature was fixed at 273 K, 
and the $\co2$ partial
pressure was varied from 1 to 37.8 bar (the saturation $\co2$ partial pressure at that temperature). A 
moist $\h2o$
adiabat was assumed in the lower troposphere, and a moist $\co2$ adiabat was used in the upper troposphere
 when condensation was encountered, following the methodology in Appendix B of \cite{Kasting1991}.

$\h2o$ and $\co2$ clouds are neglected in the model, but the effect of the former is accounted for by increasing the surface albedo, as done in previous climate simulations by the Kasting research group
\citep{Kasting1991, Jacob2008}.
 It has been argued that this 
methodology tends to overestimate the greenhouse effect of dense $\co2$ atmospheres 
\citep{GZ2011}. 
By contrast, our neglect of $\co2$ clouds may cause us to underestimate the greenhouse effect of such 
atmospheres \citep{FP1997}. 
Realistically determining the effects of clouds would require a 3-D climate model, as most clouds 
form in updrafts, which are absent in 1-D models. Some 1-D climate modeling studies include partial 
cloud coverage 
\citep{Selsis2007b} and/or parameterized microphysical cloud model \citep{CT2003,Zsom2012}, but we do not 
consider them here because we can not  model them self-consistently in our model. The effects of 
clouds on the inner and outer edge boundaries are qualitatively understood, as discussed later in the paper. 
Testing these predictions quantitatively using 3-D
climate models should be a fruitful topic for future research.

Radiative transfer was handled by methods used in recent versions of the Kasting group climate model but with updated absorption coefficients (see next section). 
A $\delta$ two-stream approximation \citep{Toon1989} was used to calculate
the net absorbed solar radiation for each of the $101$ layers, using 
separate eight-term, 
correlated-$k$ coefficients for both $\co2$ and $\h2o$ to parameterize absorption in each 
of the $38$ solar spectral intervals ranging from $0.2 - 4.5 \mu$m. These terms are convolved with each other in each spectral interval, resulting in 64 separate radiative transfer calculations per interval. 
The solar flux was averaged over six zenith angles (11.0$^{\circ}$, 25.3$^{\circ}$, 
39.6$^{\circ}$, 54$^{\circ}$, 68.4$^{\circ}$, 82.8$^{\circ}$) using Gaussian 
quadrature.
The net outgoing infrared radiation per layer was calculated using
separate eight-term correlated-$k$ coefficients for $\h2o$ and $\co2$ in 
 $55$ spectral intervals extending from $0 - 15,000$ cm$^{-1}$. 
We used double gauss quadrature in place of a standard gaussian scheme \citep{Sykes1952,TS2002}
using a code written by Ramirez.
 Half of the k-coefficients are chosen within the g-space interval 0.95-1.00 for improved resolution of the 
steeply rising portion of the cumulative distribution function, yielding smoother stratospheric temperature behavior.

These coefficients also needed to be convolved with each other, as in the solar calculation. This produces 
8$\times$ 8 $\times$ 55 $=$ 3520 separate thermal-IR
 radiative transfer calculations at each time step in the climate model. 
This number is multiplied by a factor of 6 when we include $\ch4$
in the model, using 6-term sums, and by another factor of 6 when we include C$_{2}$H$_{6}$. 
Thus, from a practical standpoint, the utility of this approach diminishes as the number of included greenhouse gases increases.

\subsection{Model Updates}
\label{modelupdate}
The following are the most significant updates to the climate model:

\begin{enumerate}

\item We have derived new {\it k-} coefficients 
 using a tool called KSPECTRUM. It is a program
to produce high-resolution spectrum of any
gas mixture, in any thermodynamical conditions, from line-by-line (LBL) databases
such as HITRAN 2008 \citep{Rothman2009} and HITEMP 2010\footnote{suggested to us by Colin 
Goldblatt, private communication} \citep{Rothman2010}.
 It is intended 
to produce reliable spectra, which can then be used to compute k-distribution data sets that 
may be used for subsequent radiative transfer analysis. The source code and a detailed 
description of the program is available at {\url {http://code.google.com/p/kspectrum/}}.

  We have produced two sets of coefficients, one using HITRAN 2008 and another using the HITEMP 2010 
database. For the HITRAN database we generated a matrix of 8-term absorption coefficients
for both $\h2o$ and $\co2$, 
using KSPECTRUM, for the following range of pressures and temperatures: 
$p(bar) = [10^{-5}, 10^{-4},10^{-3},10^{-2},10^{-1},1,10,10^{2}]$ and $T (K) = [100,150,200,
250,300,350,400,600]$. In the case of HITEMP, 8-term absorption coefficients were derived 
only for $\h2o$, as our IHZ is $\h2o$-dominated at high temperatures ($\ge 300$ K) with only trace
amounts of $\co2$ (330 parts per million). The
following grid was used to derive the $\h2o$ HITEMP coefficients:
$p(bar) = [10^{-1},1,10,10^{2}]$ and $T (K) = [350,400,600]$. The grid is condensed 
  because of the high number of line transitions in the HITEMP database compared to HITRAN. The computational resources needed to derive absorption coefficients for the entire range of pressure \& temperatures would be prohibitively large. Moreover, as 
discussed further below, we justify the selection of this condensed grid by showing
that the differences in coefficients generated from HITRAN and HITEMP become negligible below 350 K. 

In generating the $k-$coefficients, we have used different methodologies for $\co2$ and $\h2o$. 
For CO2, we truncated the spectral lines at $500$ cm$^{-1}$
from the line center. Experimental evidence indicates that the absorption by
$\co2$ is overestimated if Lorentzian line shapes are used \citep{Burch1969, Fukabori1986, 
Bezard1990, Halevy2009}. Therefore, we used the prescription of \cite{PH1989} for `sub-Lorentzian'
absorption in the far wings of the lines when running KSPECTRUM. For $\h2o$, we truncated the 
spectral lines $25$ cm$^{-1}$ and overlaid a semi-empirical ``continuum absorption''. 
The Lorentz line shape is known to underestimate absorption for $\h2o$ in the far wings
\citep{Halevy2009}, possibly because of the tendency of $\h2o$ to form dimers. The
corresponding continuum absorption is therefore `super-Lorenztian' for $\h2o$, and we have
used the `BPS' formalism of \cite{PR2011} to parameterize this absorption.

\item We have included Rayleigh scattering by water vapor, as it can become important for
wavelengths up to $1 \mu$m (which is where the Wien peak occurs for low mass stars). Rayleigh scattering 
by water was also considered by \cite{Kasting1988,Kasting1993}, but these authors used the scattering 
coefficient for air because the coefficient for $\h2o$ was not available, or at least not known to them. 
The following expression for the scattering cross\-section was adopted \citep{Allen1976, VC1984, vparis2010}:
\begin{eqnarray}
\sigma_\mathrm{R,\h2o} (\lambda)&=& 4.577 \times 10^{-21} \biggl(\frac{6+3D}{6-7D}\biggr) \biggl(\frac{r^{2}}{\lambda^{4}}\biggr)  ~~~~\mathrm{cm^{2}} 
\label{rayl}
\end{eqnarray}
Here, $D$ is the depolarization ratio ($0.17$ for $\h2o$, \cite{MS1990}),
 $r$ is the wavelength ($\lambda$)-dependent 
refractivity which is calculated as $r = 0.85 r_\mathrm{dry air}$ 
\citep{Edlen1996}, $r_\mathrm{dry air}$ is obtained from Eq.(4) of 
\cite{Bucholtz1995}, and $\lambda$ is in microns. 
By comparison, \cite{Selsis2007a} used a $\h2o$ Rayleigh scattering 
cross-section of $2.32 \times 10^{-27}$ cm$^{2}$ at $0.6~\mu$m. Evaluating Eq.(\ref{rayl}) at
$0.6 ~\mu$m gives a value of $2.6 \times 10^{-27}$ cm$^{2}$, which is similar to the 
\cite{Selsis2007a} value.
 
\item Previous climate model calculations by our group and others
 \citep{Kasting1984, Pollack1987, Kasting1991, FP1997, Mischna2000} parametrized 
collision-induced absorption (CIA) by $\co2$ near $7 ~\mu$m and beyond $20 ~\mu$m by
the formulation given in the Appendix of \cite{Kasting1984}. This process is an important source of 
thermal-IR opacity in the types of dense, $\co2$-rich atmospheres predicted to be found near the outer edge of the habitable zone.
In our model we have updated $\co2$-CIA using the parametrization described 
in \cite{GB1997, Baranov2004, Halevy2009}. 

\item The Shomate Equation\footnote{\url {http://webbook.nist.gov/cgi/cbook.cgi?ID=C124389&Units=SI&Mask=1#Thermo-Gas}} was used to calculate new heat capacity ($c_{p}$) relationships for $\co2$ and $\h2o$.
Notably, at low temperatures, the heat capacity for $\co2$ decreased by $\sim 30 \%$ relative to values in 
our previous model. This increased the dry adiabatic lapse rate, $g/c_{p}$ where $g$ is gravity, 
by an equivalent amount but had surprisingly little effect on computed surface temperatures, 
apparently because the steeper lapse rate in the upper troposphere was largely compensated by a decrease in 
tropopause height. See \cite{Ramirez2012a} for further details.

\end{enumerate}

\subsection{Model Validation}
\label{modelvalidation}
We have checked the accuracy of our climate model by comparing the output both with published results and with the 1-D line-by-line 
radiative transfer model
SMART ({\it Spectral Mapping Atmospheric Radiative Transfer}) developed by D. Crisp 
 \citep{MC1996, Crisp1997}. SMART is a well-tested model \citep{Robinson2011} which accesses 
some of the same databases as does KSPECTRUM; however, its development and implementation are entirely
independent. By comparing specific cases of interest
with SMART, we can gain confidence that our calculated fluxes are correct,
or at least that they are consistent with our assumptions about $\co2$ and $\h2o$ line shapes. 
For all our climate models that are compared with SMART, we used 70 atmospheric layers (we use
101 layers for all our HZ calculations). We could not use 101 layers
in our flux comparisons due to numerical accuracy issues with SMART at
high enough vertical resolution, although 70 layers produced a sufficiently accurate result 
with SMART.

\subsubsection{Dense $\co2$ atmosphere}
Dense $\co2$-rich atmospheres have been suggested as warming agents for early Mars \citep{Pollack1987,Kasting1991,
FP1997,Tian2010}. 
Planets close to the outer edge of the HZ may develop dense, $\co2$-rich atmospheres as
a consequence of outgassing from volcanism, which can only be balanced by surface
weathering if the planetâ€™s surface temperature remains above freezing.
The $\co2$ feedback effect
 fails at some distance because $\co2$ begins to
condense out of the atmosphere, lowering the tropospheric lapse rate and reducing the
greenhouse effect. $\co2$ is also an effective Rayleigh scatterer (2.5 times better than air),
and so a dense $\co2$ atmosphere is predicted to have a high albedo, which offsets its
greenhouse effect \citep{Kasting1991}. The OHZ boundary can then be taken as this ``maximum greenhouse limit'' where
Rayleigh scattering by $\co2$ begins to outweigh the greenhouse effect.

\begin{figure}[!hbp|t]
\includegraphics[width=0.92\textwidth]{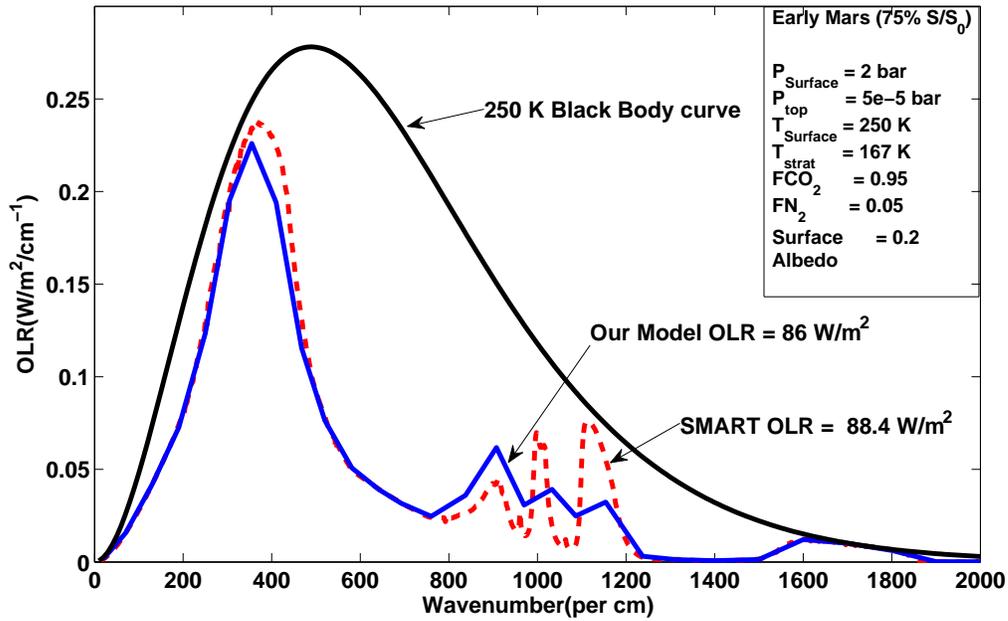}
\caption{Plot of outgoing long-wave radiation vs. wavenumber for the $0-2000$cm$^{-1}$ region
comparing our OLR (blue solid curve) to that from SMART (red dashed curve).
This calculation is for early Mars conditions, 2 bar $\co2$, and constant
stratospheric and surface temperatures of 167 and 250 K, respectively. The corresponding 250 K black body curve
is shown in black. The integrated flux over all bands at the top of the atmosphere is 
$86$ Wm$^{-2}$ for our model and $88.4$ Wm$^{-2}$ for SMART.}
\label{denseco2}
\end{figure}

Fig. \ref{denseco2} shows  net outgoing long-wave radiation (OLR) versus wavenumber in the
range $0-2000$ cm$^{-1}$ for a Mars-mass
planet with a 2-bar $\co2$ atmosphere and a surface temperature of 250 K. 
The solar constant is assumed to be $75 \%$ of its present value ($1360$ Wm$^{-2}$),
matching the solar flux incident on early Mars ($3.8$ Gyr). The integrated flux over all bands 
at the top of the atmosphere  from our
model ($86$ Wm$^{-2}$, blue solid curve) matches well with SMART
 ($88.4$ Wm$^{-2}$, dashed red curve).
Our model has a coarser spectral resolution than does SMART, and it appears that 
between
$800$ and $1200$ cm$^{-1}$ the differences could become important. But this is compensated
by the fact that our OLR in these intervals can be considered as a running average of the OLR 
 from SMART.
Nevertheless, most of the difference in the OLR arises from significant absorption in the 
$667$ cm$^{-1}$ (15 micron) vibrational band of $\co2$ which is closer to the peak of the black-body curve.

A similar study for early Mars conditions with a 2-bar $\co2$ atmosphere was considered by
\cite{Wordsworth2010}. They have also used KSPECTRUM to derive their absorption coefficients and
truncated the spectral lines at $500$ cm$^{-1}$ from the line center for $\co2$, as done here. As our surface albedo for this calculation 
(0.2) is also the same, we can directly compare the results from both studies. Fig. 2c from
\cite{Wordsworth2010} shows that the net OLR from their model is $88.17$ Wm$^{-2}$ compared to our
$86$ Wm$^{-2}$. The differences are due to the different number of atmospheric layers 
used in these models. \cite{Wordsworth2010} used 22 layers in their model, compared to 70 layers 
in our  SMART comparison climate models. The number of vertical atmospheric layers used 
in the model affects the OLR because the \cite{Toon1989} algorithm, used in both the models, 
assumes that each layer is isothermal. With few isothermal layers, more IR radiation is emitted
 from the upper part of each layer, which is a little hotter than it should be and 
which has the smallest optical depth, as measured from the top of the atmosphere\footnote{We
ran our climate model with 22 layers and found that our OLR increased to $89.1$ Wm$^{-2}$.}.

\thispagestyle{empty}

\subsubsection{Dense $\h2o$ atmosphere}
\begin{figure}[!hbp|t]
\includegraphics[width=0.92\textwidth]{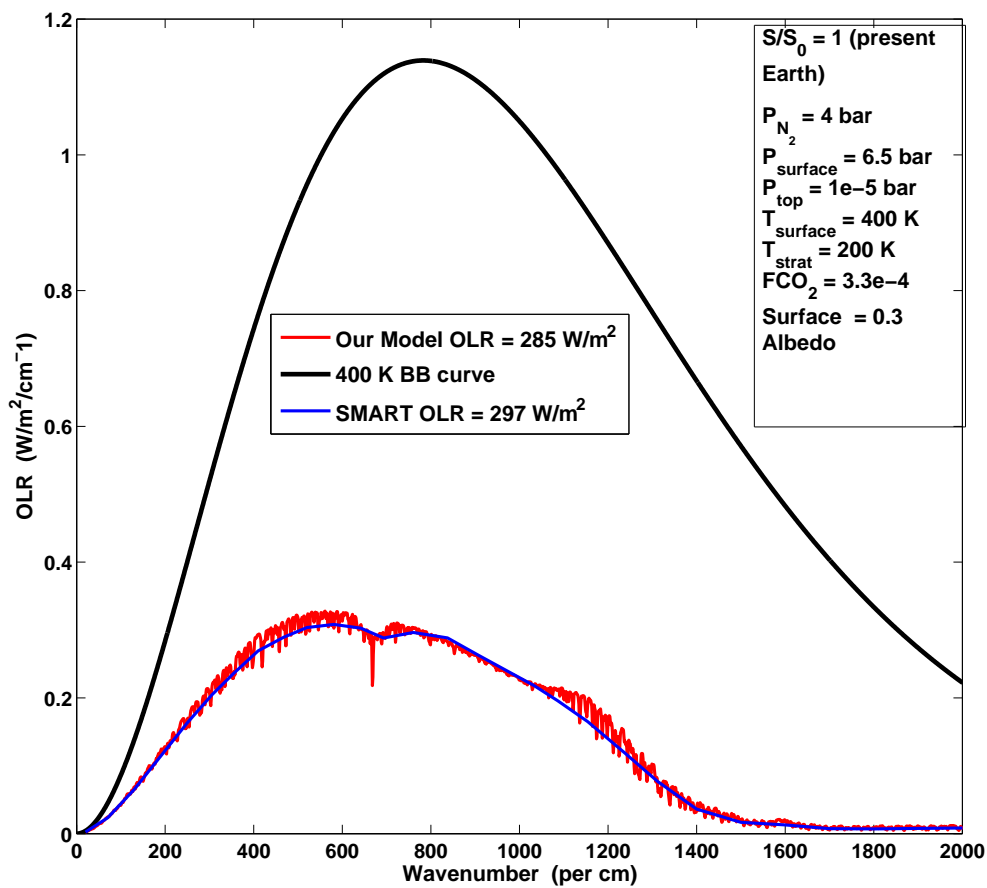}
\caption{Plot of outgoing long-wave radiation vs. wavenumber for the $0-2000$ cm$^{-1}$ region
comparing our OLR (blue solid curve) to that from SMART (red dashed curve).
This calculation is for a dense $\h2o$ atmosphere with
stratospheric and surface temperatures of 200 K and 400 K, respectively. The corresponding 400 K blackbody curve
is shown in black. Both the models appear to be in good agreement. The integrated flux over 
all bands at the top of the atmosphere is 
$285$ Wm$^{-2}$ for our model and $297$ Wm$^{-2}$ for SMART, with differences possibly arising from
different formalisms of continuum absorption (BPS versus CKD. See text for details).}
\label{denseh2o}
\end{figure}

The inner edge of the HZ in our model is determined by the
so-called ``moist greenhouse effect'', in which the stratosphere becomes water-dominated,
leading to rapid escape of hydrogen to space. Fig. \ref{denseh2o} shows the net outgoing IR 
as a function of wavenumber for a dense $\h2o$ atmosphere. Here, we assumed an Earth-mass planet
with a surface temperature of $400$ K and a surface albedo of $0.3$. The stratospheric temperature is
assumed to be constant at $200$ K. The stratosphere becomes tenuous at these high surface temperatures and has 
little effect on the outgoing IR flux.  The background gas is $4$ bar of 
N$_{2}$ and the total surface pressure is $6.5$ bar (These conditions were assumed for specific 
intercomparison with SMART for this test case). The flux incident at the top of the atmosphere is 
assumed to be the current solar flux at Earth's distance from the Sun.

As with the dense $\co2$ case, in Fig. \ref{denseh2o} we compare our model (solid blue curve) 
with SMART (dashed red curve) for the dense $\h2o$ atmosphere. 
Although both model spectra appear to be in good agreement, the
integrated flux over all bands at the top of the atmosphere from our model is $285$ Wm$^{-2}$ compared to
$297$ Wm$^{-2}$ from SMART. The differences arise in the window region of the water vapor 
($800-1200$ cm$^{-1}$) and also in between $300-600$ cm$^{-1}$, where our model absorbs more than SMART. A
possible reason is that we are using the BPS continuum, as opposed to the 'CKD' 
continuum \citep{CKD1989} used by 
SMART. The BPS formalism is based on empirical measurements which take into account the contribution of dimers, resulting in more absorption of outgoing 
IR radiation (see \cite{PR2011} Table 3; \cite{Shine2012}).

\thispagestyle{empty}

\section{Results}
\label{results}
In the subsections that follow, we estimate HZ boundaries around a star similar to our Sun.
We first compare results from our model using HITRAN and HITEMP databases, estimate HZ limits 
for non-Earth-like planets and discuss the effect of clouds on the HZ boundaries.

\subsection{Inner Edge of the HZ (IHZ)}
\label{ihz}
The inner edge of the HZ is calculated by increasing the surface temperature of a fully saturated
 ``Earth'' 
model from $220$ K up to $2200$ K. The effective
solar flux $S_{eff}$, which is the value of solar constant required to maintain a given
surface temperature, is calculated from the ratio between
the {\it net} outgoing IR flux $F_{IR}$ and the {\it net} incident solar flux $F_{SOL}$,
both evaluated at the top of the atmosphere.
The total flux incident at the top of the atmosphere is taken to be the 
present solar constant at Earth's orbit $1360$ Wm$^{-2}$. The planetary albedo is calculated as 
the ratio between the upward and downward solar fluxes.

\begin{figure}[!hbp|t]
\subfigure[] {
\label{fir}
\includegraphics[width=.50\textwidth]{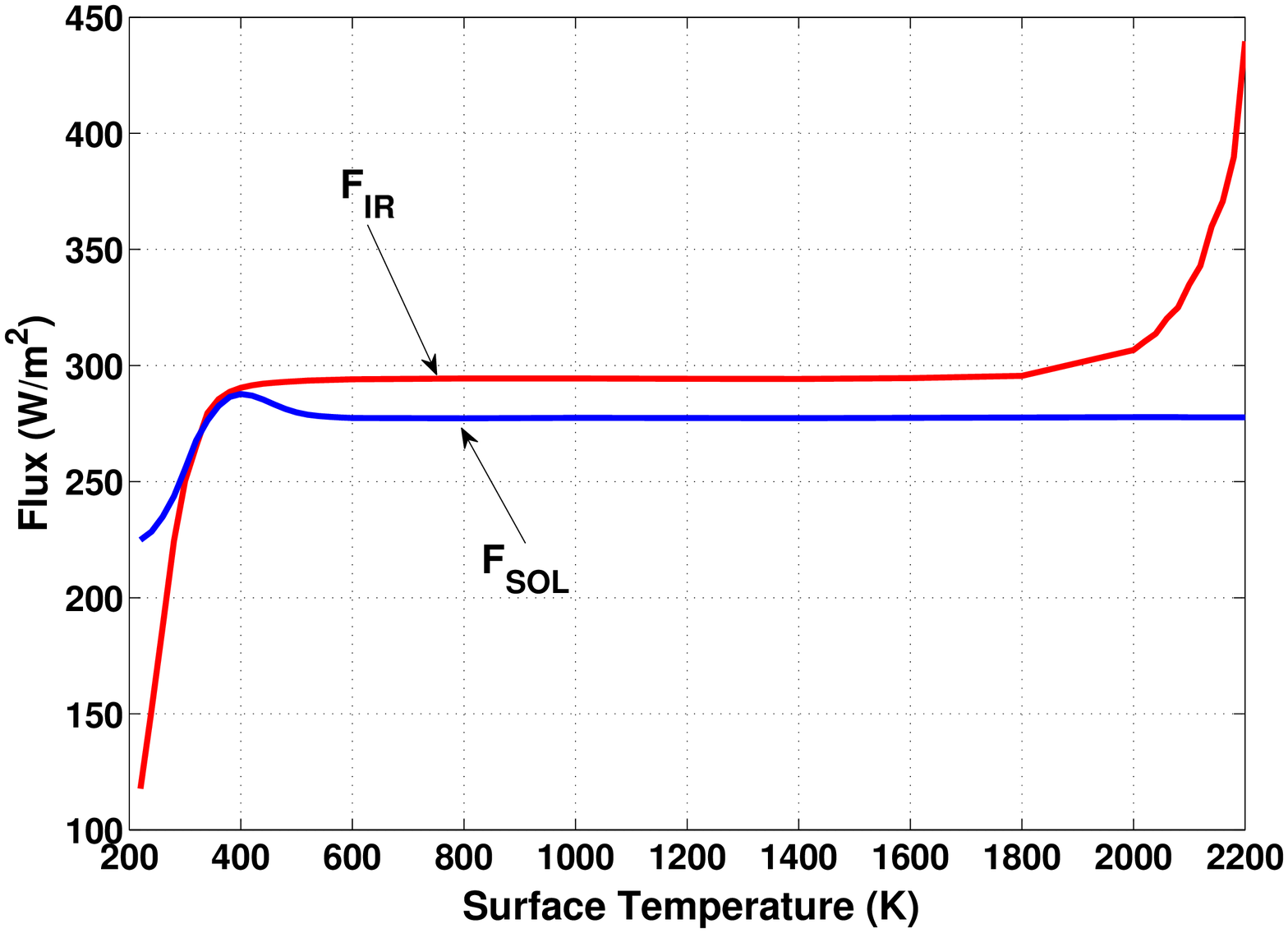}
}
\subfigure[] {
\label{albp}
\includegraphics[width=.50\textwidth]{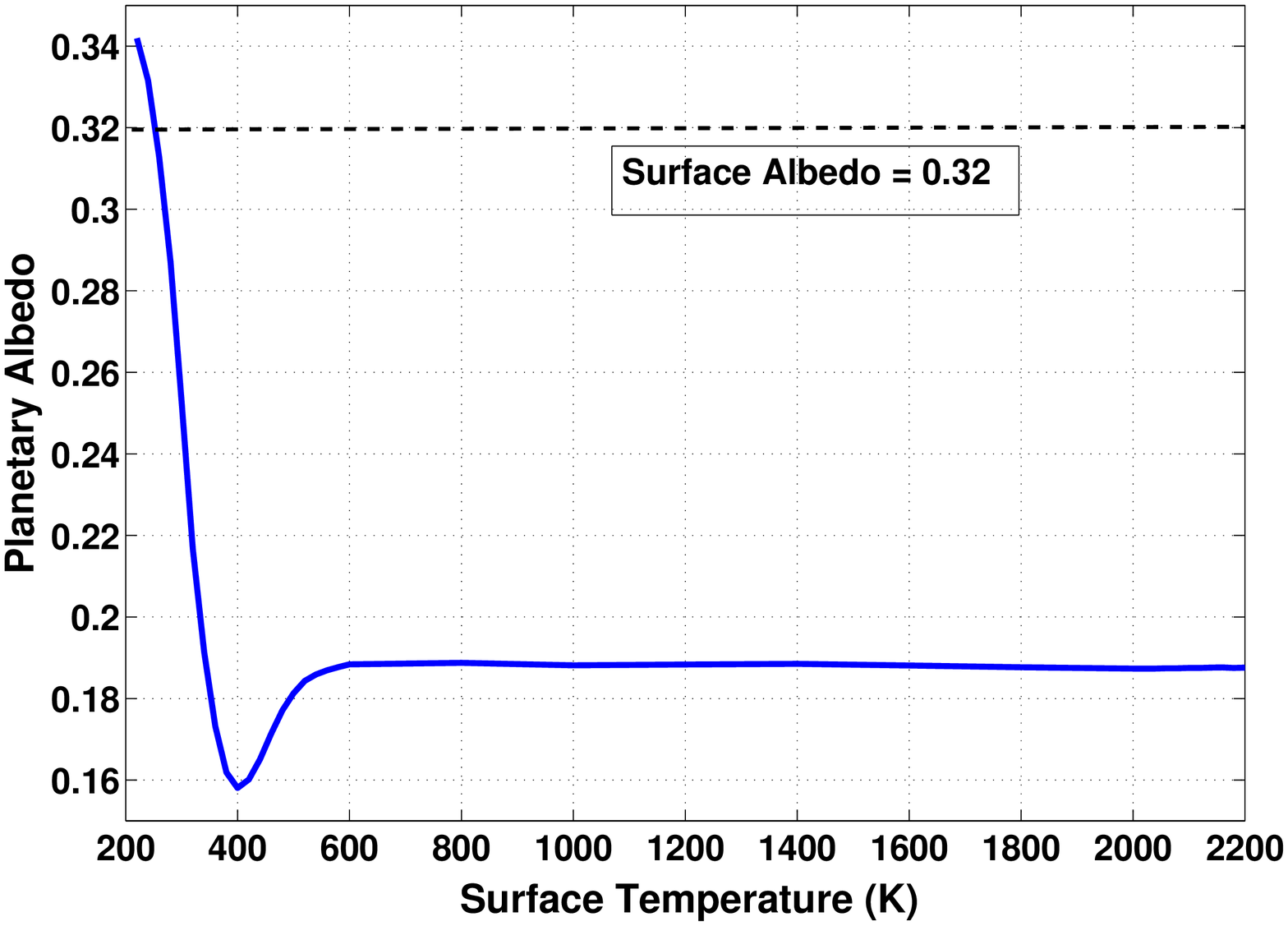}
}
\subfigure[] {
\label{seff}
\includegraphics[width=.50\textwidth]{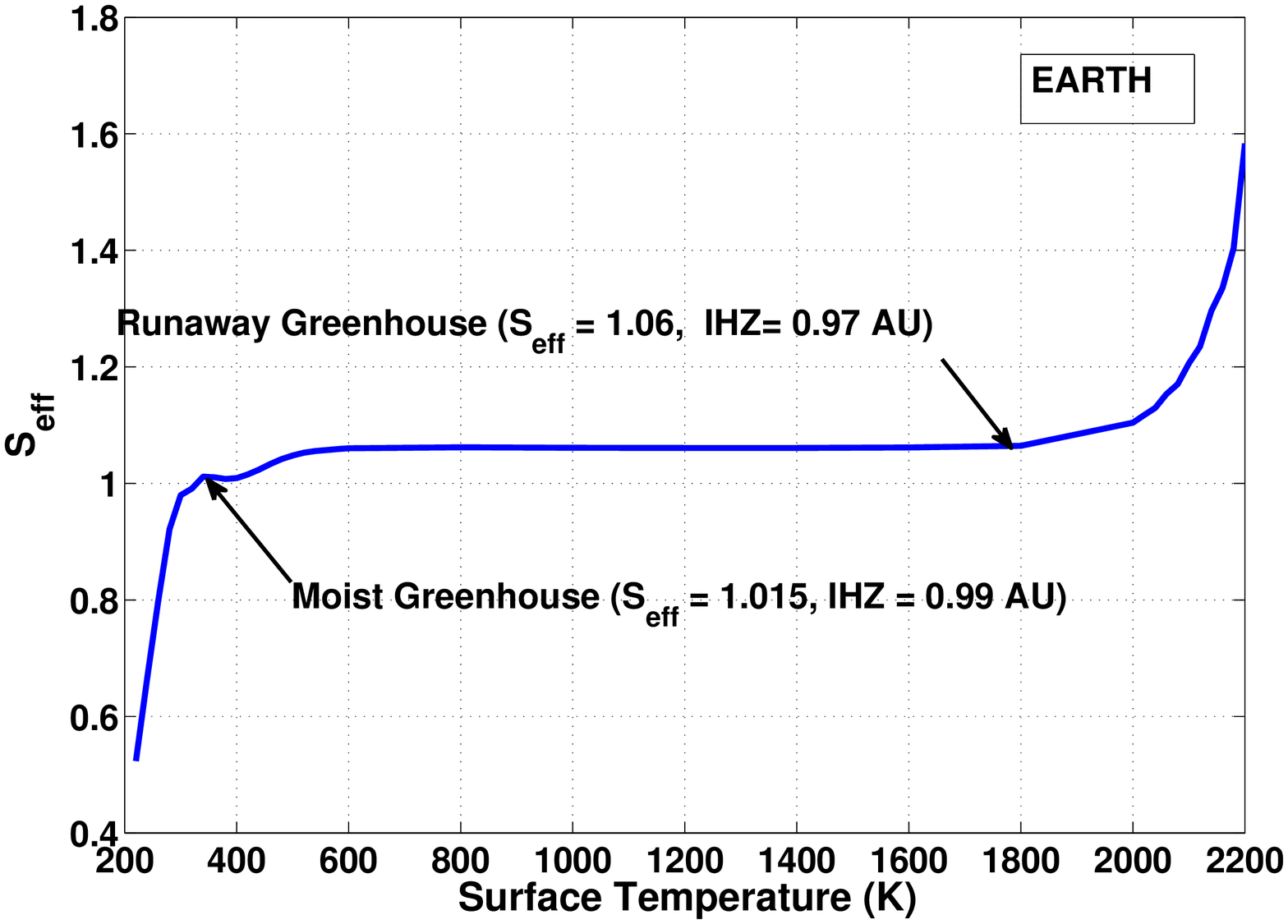}
}
\subfigure[] {
\label{fh2o}
\includegraphics[width=.50\textwidth]{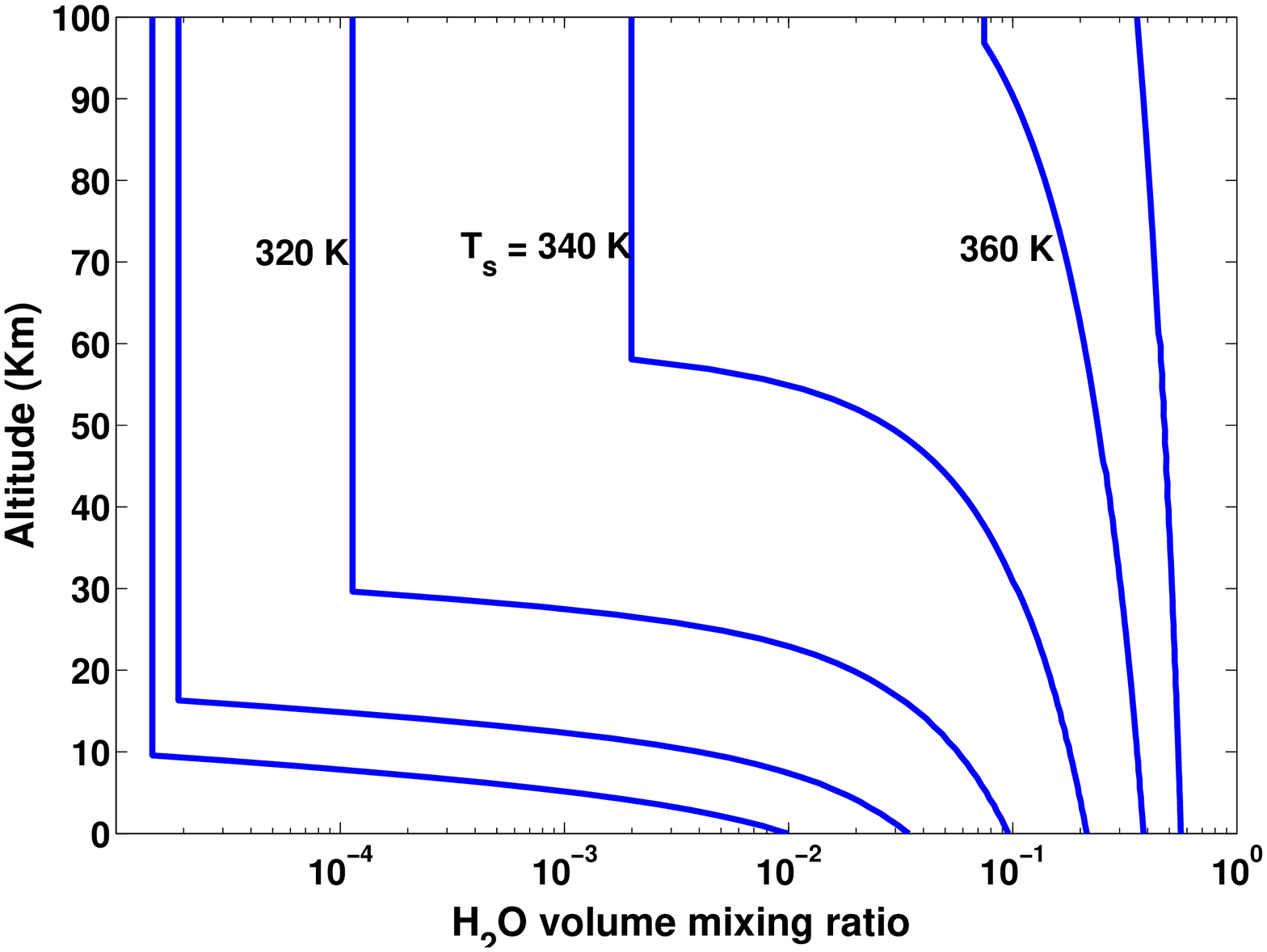}
}

\caption{Inner edge of the habitable zone calculations from our updated climate model. Various parameters are 
shown as a function of surface temperature: (a) Net outgoing IR flux and net incident solar flux (b) planetary albedo 
(c) Effective solar flux, $S_{eff} = F_{IR}/F_{SOL}$, and (d) water vapor profile. These calculations were performed with the HITEMP 2010 database. 
The water-loss (moist greenhouse) limit, which is most relevant to habitability, is at 0.99 AU and runaway greenhouse is 
at 0.97 AU. The corresponding estimates from \cite{Kasting1993} climate model are 0.95 AU for 
the moist greenhouse and 0.84 AU for the runaway greenhouse.}
\label{EarthFluxes}
\end{figure}

The calculated radiative fluxes, planetary albedo and water vapor profile for various
surface temperatures are shown in Fig. \ref{EarthFluxes}. Absorption coefficients derived
from the HITEMP 2010 database, overlaid by BPS formalism \citep{PR2011}, were used in generating these results. 
Fig. \ref{fir} shows that
$F_{IR}$ increases with surface temperature and then levels out at 291 Wm$^{-2}$, as the 
atmosphere becomes opaque to infrared radiation at all wavelengths\footnote{This value of 291 Wm$^{-2}$ closely matches
with the value from Fig. 4.37 of \cite{RayP2010} for a planet saturated with pure water vapor atmosphere and
with a surface gravity of $10$ms$^{-2}$.}. 
Beyond 2000 K, $F_{IR}$
increases again 
as the lower atmosphere and surface begin to radiate in the visible and 
near-IR, where the water vapor opacity is low. 
$F_{SOL}$ initially increases as a consequence of absorption of near-IR solar radiation by $\h2o$. It
then decreases to a constant value ($264$ Wm$^{-2}$) at higher temperatures as Rayleigh 
scattering becomes important. Planetary albedo (Fig. \ref{albp}) provides an alternative way of 
understanding this behavior.
It goes through a minimum at a surface temperature of ~400 K, corresponding to the maximum
in $F_{SOL}$, and then flattens out at a value of 0.193. 

The inner edge of the HZ for our Sun can be calculated from Fig. \ref{seff}. The behavior of 
$F_{IR}$ and $F_{SOL}$ causes $S_{eff}$ to increase initially and then remain constant at
higher temperatures. Two limits for the IHZ boundary can be calculated. The first one is
the ``moist greenhouse'' (or water-loss) limit which is encountered at a surface 
temperature of $340$ K when $S_{eff} = 1.015$. At this limit, the water vapor content in the
stratosphere increases dramatically, by more than an order of magnitude, as shown in 
Fig. \ref{fh2o}. This is the relevant
IHZ boundary for habitability considerations, although it should be remembered that the actual inner edge may be closer to the Sun if cloud feedback tends to cool the planet's surface, as expected.\footnote{The total $\h2o$ inventory assumed here is equal
to the amount of water in Earth's oceans -- $1.4 \times 10^{24}$ grams. This amounts to 
$2 \times 10^{28}$ atoms per cm$^{-2}$. Once the stratosphere becomes wet, water vapor photolysis releases
hydrogen which can escape to space by diffusion limited escape rate. The time scale for water loss approaches the age
of the Earth when the mixing ratio of water is $\sim 3 \times 10^{-3}$, which happens at a surface temperature of 
340 K.} The orbital distance 
corresponding to the cloud-free water loss limit is  
$d= 1/S_{eff}^{0.5}=0.99$ AU for an  Earth-like planet orbiting the Sun.

The second IHZ limit is the runaway greenhouse at which the oceans evaporate entirely. The limiting $S_{eff}$
from Fig. \ref{seff} is 1.06 which corresponds to a distance of 0.97 AU. 
Both calculated IHZ limits are significantly
farther from the Sun than the values found by \cite{Kasting1993} ($0.95$ AU for the water-loss limit and $0.84$ AU for the
runaway greenhouse). The difference is caused by increased atmospheric absorption of incoming solar 
radiation by $\h2o$ in the new model. As pointed out by \cite{Kasting1993}, 
a third estimate for the IHZ boundary can be obtained from radar 
observations of Venus by Magellan spacecraft, which suggest that liquid water has been absent from the 
surface of Venus for at least 1 Gyr \citep{SH1991}. The Sun at that time was $\sim92 \%$ of the present day
luminosity, according to standard stellar evolutionary models \citep[See Table 2]{Baraffe1998, Bahcall2001}. 
The current solar flux at Venus distance is $1.92$ times that of Earth. 
 Therefore, the solar flux received by Venus at that time was $0.92 \times 1.92 = 1.76$ times that of Earth.
This empirical estimate of the IHZ edge corresponds to an
orbital distance of $d=(1/1.76)^{0.5} = 0.75$ AU for the present day. Note that this distance
is greater than Venus' orbital distance of 0.72 AU because the constraint of surface water was imposed at an earlier time in the planet's history.

\thispagestyle{empty}

\subsection{Comparison of inner edge results using the HITEMP and HITRAN databases}
\begin{figure}[!hbp|t]
\includegraphics[width=0.92\textwidth]{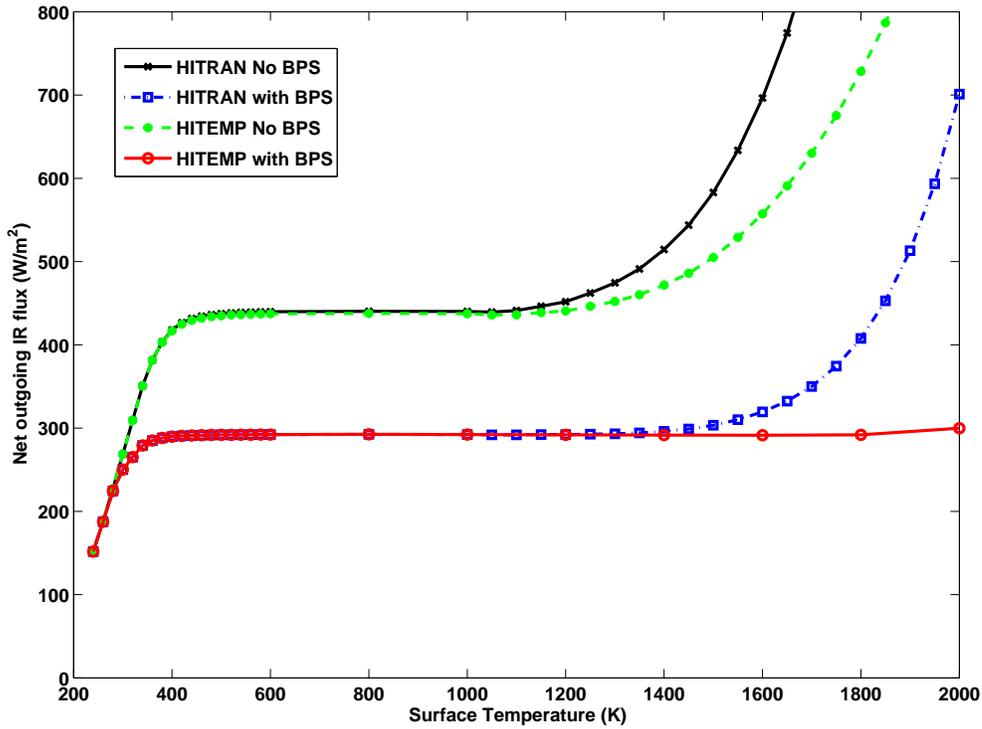}
\caption{Comparison of outgoing IR radiation ($F_{IR}$) from HITEMP and HITRAN databases,
with (blue \& red curves) and without (black \& green curves) continuum absorption.
$F_{IR}$ is lower in the 'with BPS' case for both the databases because the continuum 
absorption becomes significant in the water vapor window regions.}
\label{BPS}
\end{figure}

In Fig. \ref{BPS} we show $F_{IR}$ as a function of surface temperature (similar to Fig. \ref{fir}).
We wish to compare the outgoing IR calculated from HITRAN \& HITEMP databases with and without
overlaying the continuum absorption. Fig. \ref{BPS} shows two significant differences: 
\begin{enumerate}

\item The limiting value of $F_{IR}$ which leads to a runaway greenhouse happens at a much higher value
(440 Wm$^{-2}$, black \& green curves) when the BPS $\h2o$ continuum formalism is not implemented, 
and at a lower $F_{IR}$ (291  Wm$^{-2}$, red \& blue curves) 
when the BPS continuum is included in our model.  The continuum is based on 
measurements of absorption in the water vapor window regions (i.e $800 - 1200$ cm$^{-1}$  and
$2000 - 3000$ cm$^{-1}$).
 At high temperatures, the contribution of the continuum absorption in these window 
regions becomes significant, and this, in turn, decreases the outgoing IR flux.

\item The moist-greenhouse (water loss) limit moves much closer to the Sun (to 0.87 AU) when
continuum absorption is not included, as compared to 0.99 AU when it is included in our model.
This is a direct consequence of the differences in $F_{IR}$ described above: When $F_{IR}$
increases with the continuum turned off, $S_{eff}$ (ratio of $F_{IR}$ to $F_{SOL}$) 
increases and the IHZ distance  $d= 1/S_{eff}^{0.5}$ decreases. The result can
be understood physically by noting that in the model where the continuum is absent, 
the planet needs more effective solar flux to maintain a given surface temperature because
more thermal-IR radiation leaks away into space; hence, the IHZ boundary must move inward. 

A similar change can be seen in the runaway greenhouse limit: the `No BPS' model transitions
to runaway at a higher $S_{eff}$ than does the `with BPS' model. The corresponding runaway greenhouse
limit changes from 0.97 AU (with continuum absorption) to 0.76 AU (without continuum absorption).
Fig. \ref{BPS} also shows that the upturn in $F_{IR}$ beyond 800 K happens at lower surface
temperatures when continuum absorption is not included. It should be remembered that this upturn
in $F_{IR}$ happens because, 
as the surface warms, the region in the troposphere over which the temperature profile follows a dry adiabat 
expands upward, while the moist convective layer in the upper troposphere becomes thinner. Eventually, when the 
moist convective region (the cloud layer) begins thin enough, radiation emitted from the dry adiabatic 
portion of the atmosphere begins to escape to space.
The dry adiabatic lapse rate is steeper than the moist adiabatic lapse rate by about a factor of 
9 ($\sim 10$ K/km vs. 1.1 K/km); hence, the emitted radiation flux is much higher. This can be understood from the 
integrated form of Schwarzchild's equation, which shows that the emitted flux is proportional to the temperature gradient. 
(See, e.g., eq. A4 in \cite{Kasting1984}) Unlike \cite{Kasting1988}, we find that the emitted flux increases at all 
thermal-IR wavelengths shorter than 4 $\mu$m. The amount of visible radiation emitted remains negligible for surface 
temperatures of 2200 K or below.
Without the continuum 
there are fewer lines to cause absorption in these thermal-IR bands and hence 
a lower temperature would suffice to cause the upturn. Fig. \ref{BPS} also shows that the model
that includes both HITEMP \& continuum (red curve) is the one that absorbs the most outgoing
IR radiation (which is the one that was used to derive inner HZ limits in \S\ref{ihz}). 
\end{enumerate}

\subsection{Outer Edge of the HZ  (OHZ)}
\label{ohz}
\begin{figure}[!hbp|t]
\subfigure[] {
\label{firco2}
\includegraphics[width=.50\textwidth]{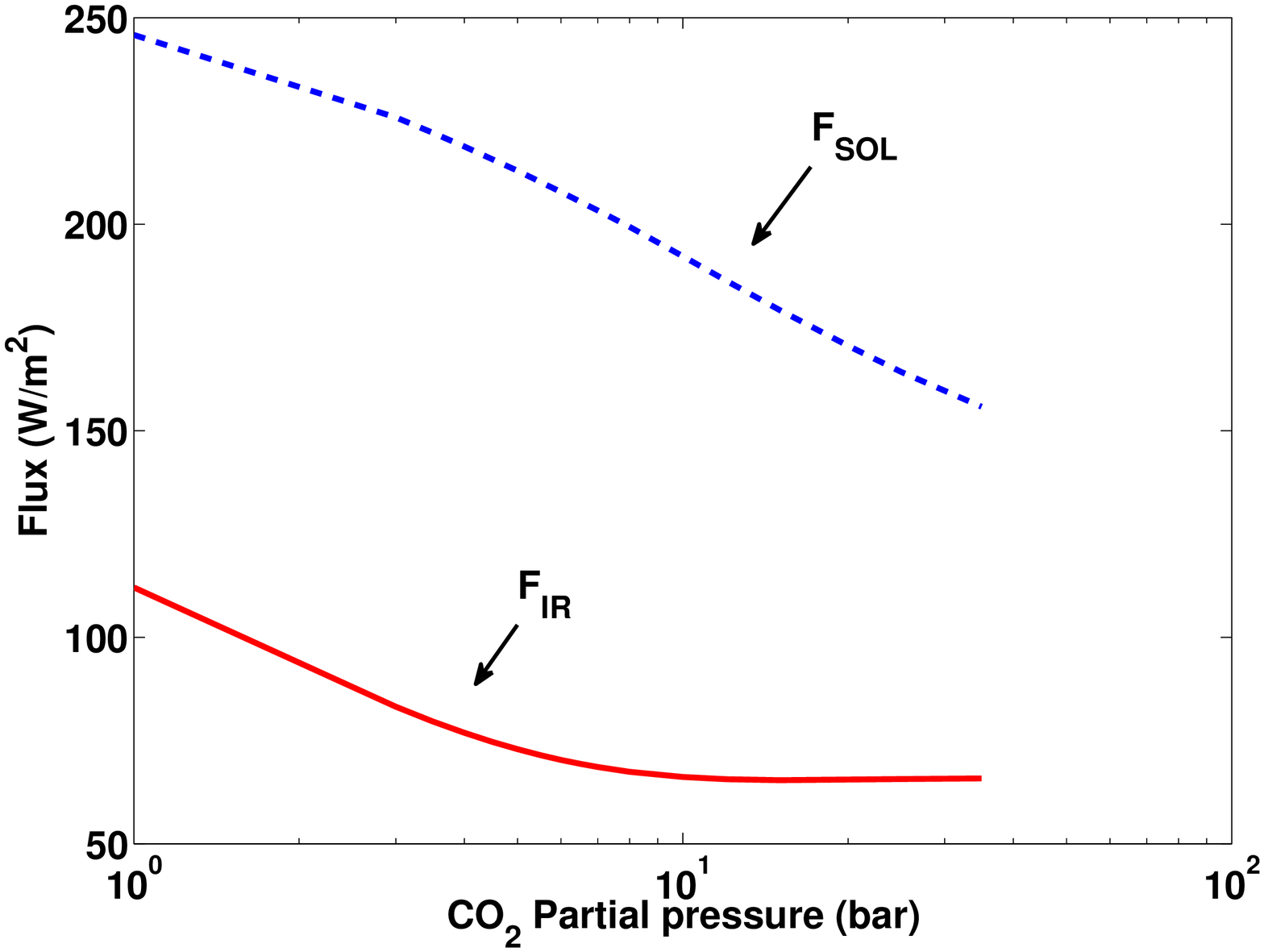}
}
\subfigure[] {
\label{albpco2}
\includegraphics[width=.50\textwidth]{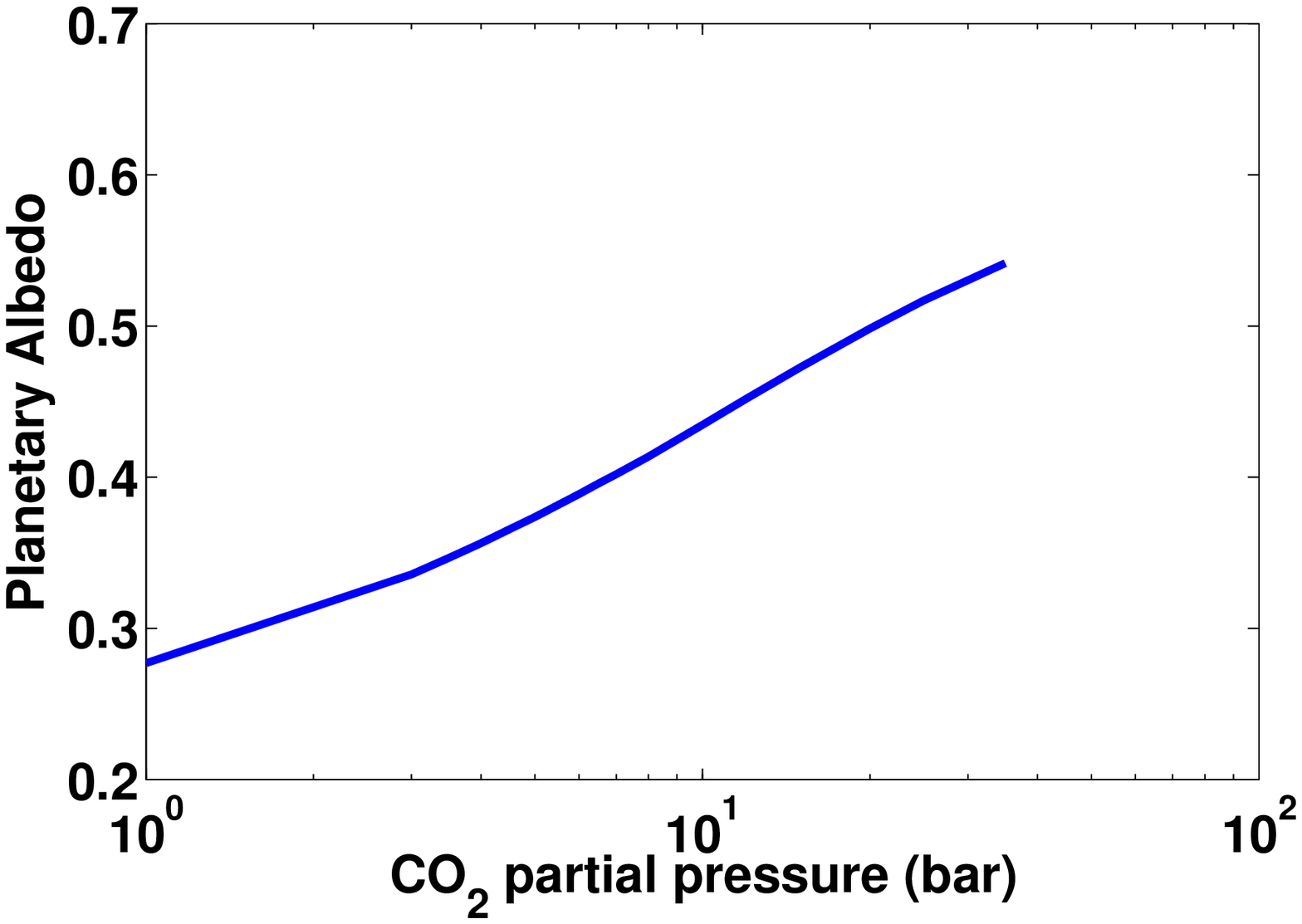}
}
\begin{center}
\subfigure[] {
\label{seffco2}
\includegraphics[width=.50\textwidth]{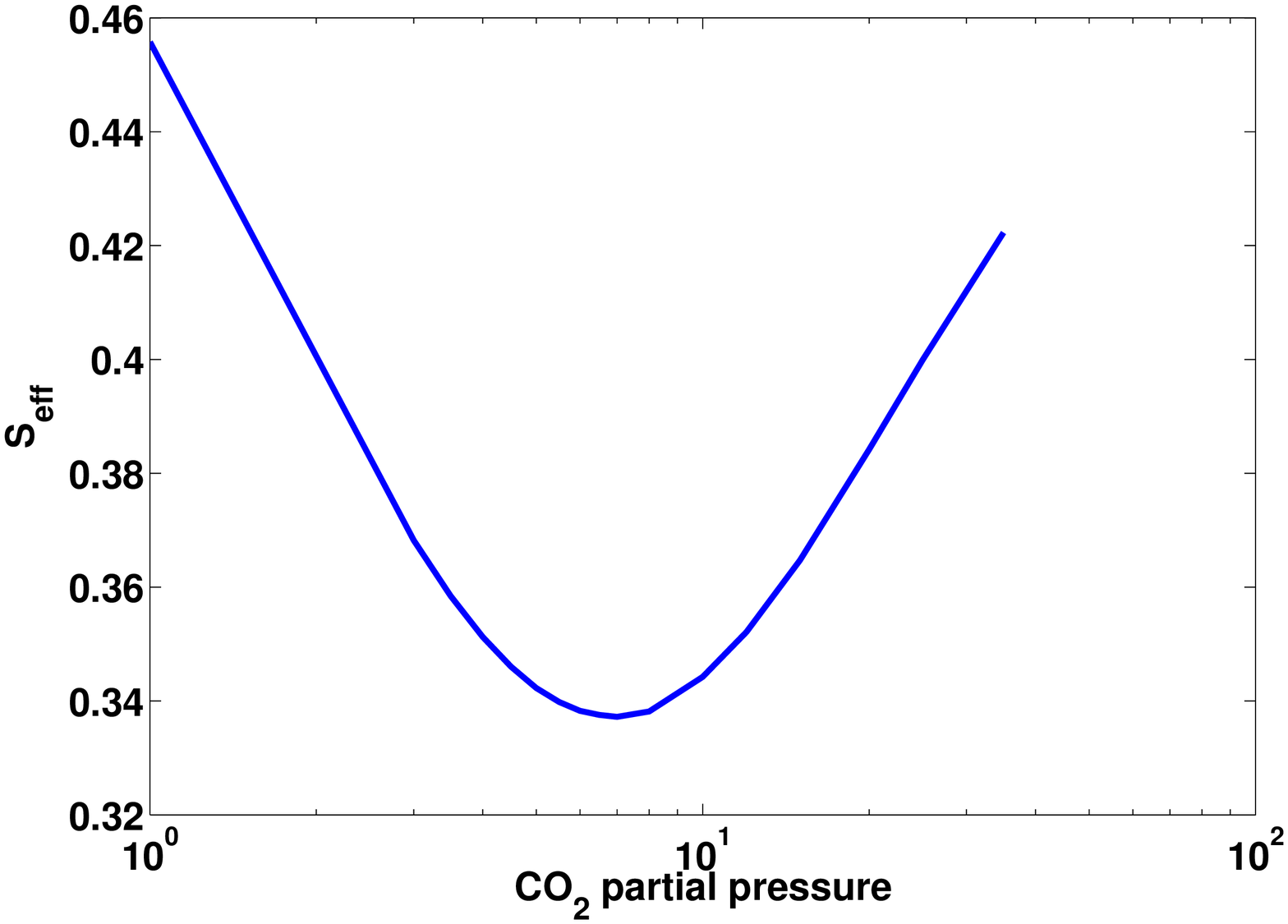}
}
\end{center}

\caption{Outer edge of the habitable zone calculations from our climate model, 
shown as a function of $\co2$ partial pressure $p\co2$: (a) Net outgoing IR flux and net incident solar flux (b) planetary albedo 
(c) Effective solar flux. 
The maximum greenhouse limit, where the atmosphere becomes opaque to outgoing IR radiation,
is at 1.70 AU ($S_{eff} = 0.343$). The previous estimate from \cite{Kasting1993} climate model was 1.67 AU.}
\label{outerfluxes}
\end{figure}

In determining the outer edge of the HZ, the surface temperature of an Earth-like planet 
with 1-bar N$_{2}$ atmosphere was fixed at 273 K and the atmospheric $\co2$ partial pressure, $p\co2$,
was varied from 1 to 35 bars (the saturation vapor pressure for $\co2$ at that temperature).
The stratospheric temperature was chosen as follows: The  model atmosphere (Mars-like 
planet) in which the onset of $\co2$ condensation occurs has a cold-trap temperature of 
154 K at an altitude where the ratio of the saturation vapor pressure to the ambient
pressure is unity. We replace the temperature profile above this altitude with a constant
temperature of 154 K. 
This allows us to calculate the solar flux ($S_{eff}$) required to maintain a global 
mean surface temperature of 273 K 
as explained in \S\ref{ihz}.
Our working hypothesis is that atmospheric $\co2$ would accumulate as these planets cooled 
because of the negative feedback provided by the carbonate-silicate cycle. Results from
our model calculations are shown in Fig. \ref{outerfluxes}.

The incident solar ($F_{SOL}$) and outgoing IR ($F_{IR}$) fluxes are shown in 
Fig. \ref{firco2}. $F_{IR}$ decreases initially as $\co2$ partial pressure is increased; this 
is an indication of greenhouse effect of $\co2$. At $\sim 10$ bars, $F_{IR}$ 
asymptotically approaches a constant value as the atmosphere
becomes optically thick at all infrared wavelengths. $F_{SOL}$ decreases monotonically with
increases in $\co2$ partial pressure as a result of increased Rayleigh scattering. 
Correspondingly, the planetary albedo increases to high values at large
 $\co2$ partial pressures, as shown in Fig. \ref{albpco2}.
The solar and IR fluxes, acting in opposite directions, create a minimum of $S_{eff} = 0.325$
 at a $\co2$ partial pressure of $\sim 8$ bar (Fig. \ref{seffco2}),
 corresponding to a distance $d = 1.70$ AU.
This defines the maximum greenhouse limit on the outer edge of the HZ. By comparison,
\cite{Kasting1993} model predicted $d = 1.67$ AU for the maximum greenhouse limit.
 As emphasized earlier, 
radiative warming by $\co2$ clouds is neglected here, even though they should be present in this 
calculation. Therefore, our OHZ limit should be considered as a conservative estimate, that is, the real outer edge is probably farther out.

As with the inner edge model, a more optimistic empirical limit on the OHZ can be estimated based on the 
observation that
early Mars was warm enough for liquid water to flow on its surface \citep{Pollack1987, Bibring2006}.
Assuming the dried up riverbeds and valley networks on martian surface are 3.8 Gyr old, the solar luminosity at
that time would have been $\sim 75 \%$ of the present value (See Eq.(1) in \cite{Gough1981} and Table 2
in \cite{Bahcall2001}). The present-day solar flux at Mars distance is $0.43$ times that of Earth. Therefore,
 the solar flux received by Mars at 3.8 Gyr was $0.75 \times 0.43 = 0.32$ times that of Earth.
 The corresponding OHZ limit today, then, would be
$d = (1/0.32)^{0.5} \approx 1.77$ AU.

Note that this distance exceeds the maximum greenhouse limit of 1.70 AU
estimated above indicating that to keep early Mars wet, additional greenhouse gases other $\co2$ and 
$\h2o$ may be required. In fact, \cite{Ramirez2012a} show that a 3-bar atmosphere 
containing 90 percent $\co2$ and 10 percent H$_{2}$ could have raised the mean
surface temperature of early Mars above the freezing point of water. The warming is caused by the
collision-induced absorption due to foreign-broadening by molecular hydrogen.
It should be acknowledged that some authors 
(e.g., \cite{Segura2002, Segura2008}) do not agree that early Mars must have been warm; however, in our view, 
these cold early Mars models do not produce enough rainfall to explain valley formation \citep{Ramirez2012a}.

\thispagestyle{empty}

\subsection{Effect of clouds on the HZ boundaries}
\label{clouds}
We summarize various cloud-free HZ boundary estimates for Earth in Table \ref{table1}.
Although we updated our radiative transfer model to incorporate new
absorption coefficients, this by itself may not yield a significantly better estimate for the
width of the habitable zone. The reason is that it is widely acknowledged that the HZ
boundaries will be strongly influenced by the presence of clouds. $\h2o$ clouds should
move the inner edge inwards \citep{Kasting1988, Selsis2007b} because their
contribution to a planet's albedo is expected to outweigh their contribution to the
greenhouse effect. (A dense $\h2o$ atmosphere is already optically thick throughout most
of the thermal-IR, so adding clouds has only a small effect on the outgoing IR radiation.)
Conversely, $\co2$ ice clouds are expected to cause warming in a dense CO2 atmosphere
because they backscatter outgoing thermal-IR radiation more efficiently than they
backscatter incoming visible/near-IR radiation \citep{FP1997}. One
can demonstrate the nature of these cloud influences using 1-D models, as was done in
\cite{Selsis2007b}. Making quantitative statements is difficult, however, because the
warming or cooling effect of clouds depends on a host of parameters, including their
heights, optical depths, particle sizes, and most importantly, fractional cloud coverage.
\cite{FP1997} obtained as much as 70 degrees of warming out of an optical
depth 10 $\co2$ cloud with $100\%$ cloud cover, but that warming dropped by 30 degrees if
fractional cloud cover was reduced to $75 \%$. Realistic fractional cloud cover for
condensation clouds is closer to $50\%$, because such clouds tend to form on updrafts,
and approximately half the air in the troposphere is rising at any one time while the other half is descending.

The best way to incorporate cloud effects in a climate calculation is to use a 3-D
general circulation model (GCM). Attempts were made to explain warm early mars using such 3-D  models 
\citep{Forget2012} but none have yet succeeded. 
One can, however, do
significantly better than in our 1-D model, and so further research in this area is warranted 
\citep{Abe2011,Wordsworth2011}.

\begin{table}[h!]
\caption{Habitable Zone distances around our Sun from our updated 1-D climate model. For comparison, estimates
from \cite{Kasting1993} are also shown. }
\vspace{0.1 in}
\begin{tabular}{|c|c|c|c||c|c|}
\hline
\multicolumn{6}{|c|}{~~~~~~~~~~~~~~~~~~~~~~~~~~~~~~~~~~~Inner Habitable Zone ~~~~~~~~~~~~~~~~~~~~~~Outer Habitable
Zone} \\
\hline
Model & Moist & Runaway  & Recent Venus & Maximum& Early Mars\\
& greenhouse & greenhouse && greenhouse &\\
\hline
This paper & 0.99 AU & 0.97 AU &0.75 AU & 1.70 AU & 1.77 AU\\
&&&&&\\
\cite{Kasting1993} & 0.95 AU & 0.84 AU & 0.75 AU & 1.67 AU & 1.77 AU\\
\hline
\end{tabular}
\label{table1}
\end{table}

\subsection{Habitable Zone Limits for Non-Earth-like planets}
In Table \ref{table2}, we show the effect of surface gravities on the HZs of two planets.
These planetary gravities were selected to encompass the mass range from Mars (gravity of 3.73 
ms$^{-2}$) to a roughly 10 M$_{\oplus}$ super-Earth (gravity of 25 ms$^{-2}$). Both planets were 
assumed to have a 1 bar
background N$_{2}$ atmosphere. This may be unrealistic because proportionately more nitrogen is
put on the smaller planet than the larger one; however, this allows direct comparison with 
\cite{Kasting1993}. Table \ref{table2} shows that the habitability 
limits move slightly outward for a Mars-sized planet and inward for a super-Earth. 
This is because the column depth is larger for a Mars-sized planet, which increases
the greenhouse effect (at the inner edge) and albedo (at the outer edge). Since the inner edge
moves closer to the star for the super-Earth planet, while the outer edge changed little, we 
can conclude that, for a given surface pressure, larger planets have somewhat wider 
habitable zones than do small ones.

\begin{threeparttable}[h!]
\caption{Habitable Zones around our Sun for different planetary parameters.}
\vspace{0.1 in}
\centering
\begin{tabular}{|c|c|c||c|}
\hline
\multicolumn{4}{|c|}{~~~~~~~~~~~~~~~~~~~~~~~~~~Inner Habitable Zone ~~~~~~~~~~~~Outer Habitable
Zone} \\
\hline
Model & Moist & Runaway  &  Maximum\\
& greenhouse & greenhouse & greenhouse\\
\hline
Mars-sized planet$^{\ast}$ & 1.035 AU & 1.033 AU &1.72 AU\\
&&&\\
Earth & 0.99 AU & 0.97 AU &1.70 AU\\
&&&\\
Super-Earth$^{\ast \ast}$ & 0.94 AU & 0.92 AU &1.67 AU\\
&&&\\
 $ p  \co2 = 5.2 \times 10^{-3}$  bar$^{\dagger}$ & {\bf 1.00 AU} & {\bf 0.97 AU} & --\\
&&&\\
$p\co2 = 5.2 \times 10^{-2}$ bar & 1.02 AU & 0.97 AU & --\\
&&&\\
$p\co2 = 5.2 \times 10^{-1}$ bar & 1.02 AU & 0.97 AU & --\\
&&&\\
$p\co2 = 5.2 $ bar & 0.99 AU & 0.97 AU & --\\
\hline
\end{tabular}
\begin{tablenotes}
\begin{footnotesize}
\item $^{\ast}$ Surface gravity = 3.73 m.s$^{-2}$
\item $^{\ast \ast}$ Surface gravity = 25 m.s$^{-2}$
\item $^{\dagger}$ $p\co2 = 5.2 \times 10^{-4}$ bar for our standard Earth model. Note that these $\co2$
 pressures are not actual partial pressures; rather, they represent the surface pressure that would be 
produced if this amount of $\co2$ were placed in the atmosphere by itself. The 330 ppmv of $\co2$ in our 
standard 1-bar atmosphere would produce a surface pressure of $5.2  \times 10^{-4}$ bar if the rest of the 
atmosphere was not present. When lighter gases such as N$_{2}$ and O$_{2}$ are present, they increase the 
atmospheric scale height and cause $\co2$ to diffuse upward, thereby lowering its partial pressure at 
the surface.
\end{footnotesize}
\end{tablenotes}
\label{table2}
\end{threeparttable}

We also performed sensitivity tests on the inner edge of the HZ by varying the amount of 
atmospheric $\co2$ (the outer edge calculation already factors in this change in $\co2$). 
It is quite possible that some terrestrial planets may have varying amount of $\co2$ because of 
different silicate weathering rates. As shown in Table. \ref{table2}, changes in
$p\co2$ would not change the runaway greenhouse limit, as it is reached in an $\h2o$-dominated
atmosphere. The moist greenhouse limit does change, as an increase in $p\co2$ increases the 
surface temperature, and hence facilitates water loss. The maximum destabilization occurs
at a $p\co2 = 5.2 \times 10^{-3}$ bar approximately 10 times the present 
terrestrial $p\co2$ level 
(the critical distance, shown in bold in Table \ref{table2}, is 1.00 AU).

This suggests that a
10-fold increase in $\co2$ concentration relative to today could push Earth into a moist greenhouse state (assuming a fully saturated atmosphere). 
By contrast, the maximum destabilization occurred at $1000$ times the present $\co2$ level in \cite{Kasting1993}. At larger
$p\co2$ values the increase in surface pressure outstrips the increase in the saturation
vapor pressure of water, so the atmosphere becomes more stable against water loss 
\citep{KA1986}. 
We conclude that planets with few tenths of a bar of $p\co2$ have narrower HZs than
planets like Earth on which $p\co2$ is maintained at lower values by the carbonate-silicate cycle.

\section{Habitable Zones around Main-Sequence Stars}
\label{FGKM}
The procedure described in the previous section to derive HZs
around Sun can be used to estimate HZ boundaries around stars of different spectral
types. A similar analysis was done by \cite{Kasting1993} for three stellar effective temperatures
(7200 K, 5700 K and 3700 K), which correspond to F0, G0, and M0 spectral types. 
\cite{Selsis2007b} used a similar model to that of \cite{Kasting1993} and interpolated HZ distances 
to stars within this range of effective temperatures. 
Here, we compare our updated model results with these earlier studies and also extend the calculations  to lower stellar effective temperatures to include M-dwarfs.
Correctly calculating HZs of M-dwarfs is becoming increasingly
important, as upcoming instruments such as Penn State's stabilized fiber-fed 
near-infrared (NIR) spectrograph {\it Habitable Zone Planet Finder}
(HPF, \cite{Suvrath2012})
and proposed missions such as {\it Transiting Exoplanet Survey Satellite} (TESS)
will specifically search for low-mass planets around M-dwarfs. Furthermore, several 
rocky planets have already been found in the HZs of M-dwarfs
\citep{Bonfils2011, Vogt2012}, and these objects may be good candidates for space-based 
characterization missions such as {\it JWST}.

\subsection{Habitable Zone Boundaries Around F, G, K and M Stars}

We considered stellar effective temperatures in the range $2600$ K $ \le T_{eff} \le 7200$ K, which 
encompasses F, G, K and M main-sequence spectral types. As input spectra for the HZ
boundary calculations we used the ``BT$\_$Settl'' grid of models\footnote{\url{http://perso.ens-lyon.fr/france.allard/}} \citep{Allard2003, Allard2007}.
These cover the needed wavelength range for climate models (0.23-4.54
$\mu$m), as well as the range of effective temperature ($2600$ K $\le T_{eff} \le 70,000$ K) 
needed to simulate stellar spectra.
Our comparison of the BT$\_$Settl models with low-resolution IRTF data, 
and also high-resolution CRIRES data on
Barnard's star (from the CRIRES$\_$POP library\footnote{\url{http://www.univie.ac.at/crirespop/}}  \citep{Lebzelter2012}), 
show that the models are quite good in
reproducing the gross spectral features and energy distributions of stars, and will provide adequate input for our HZ calculations. 
For each star, the total energy flux over our climate
model's spectral bands is normalized to 1360 Wm$^{-2}$ (the present solar constant for Earth)
to simplify intercomparison.

\begin{figure}[!hbp|t]
\subfigure[] {
\label{albtg0}
\includegraphics[width=.50\textwidth]{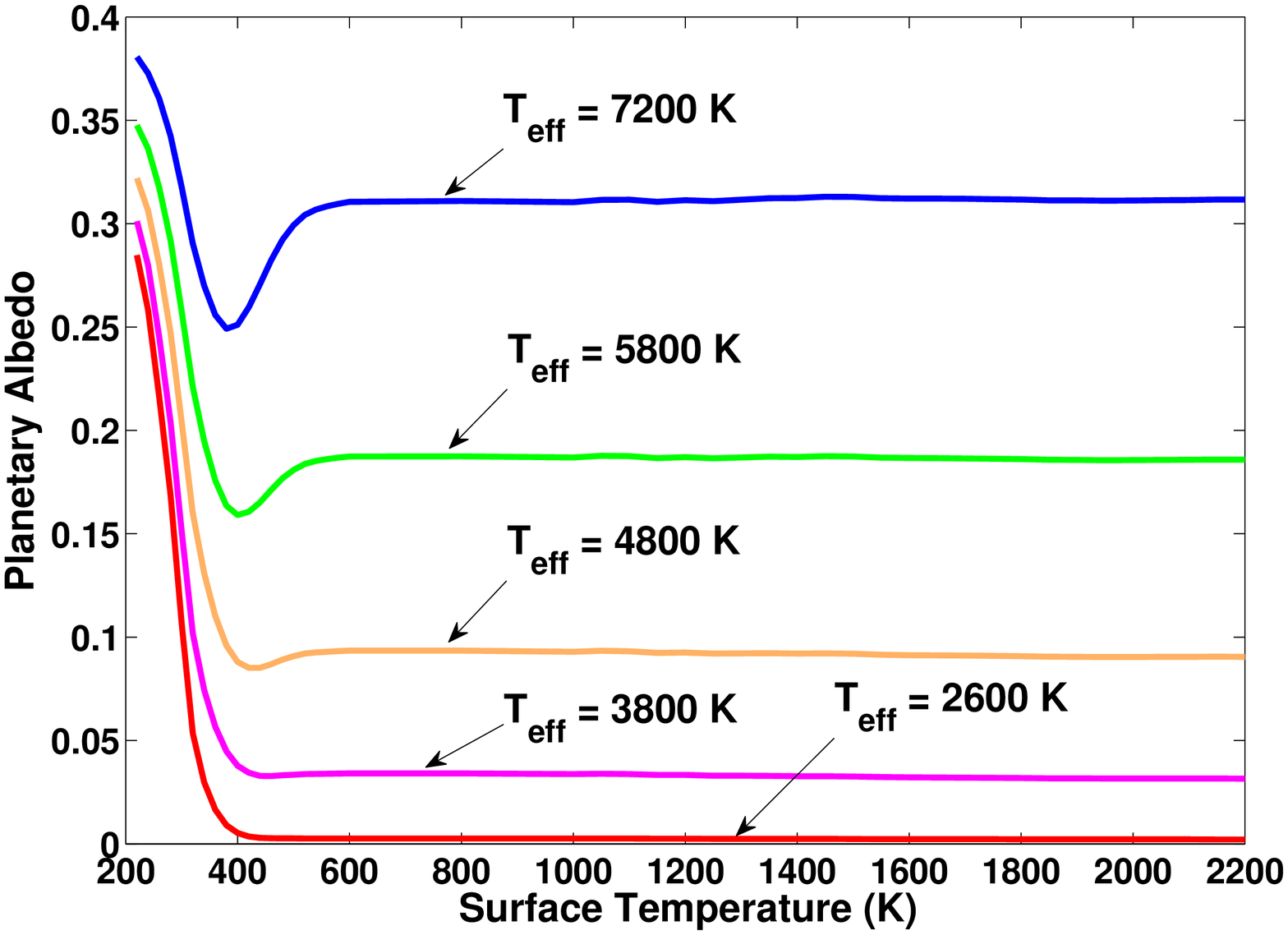}
}
\subfigure[] {
\label{sefftg0}
\includegraphics[width=.50\textwidth]{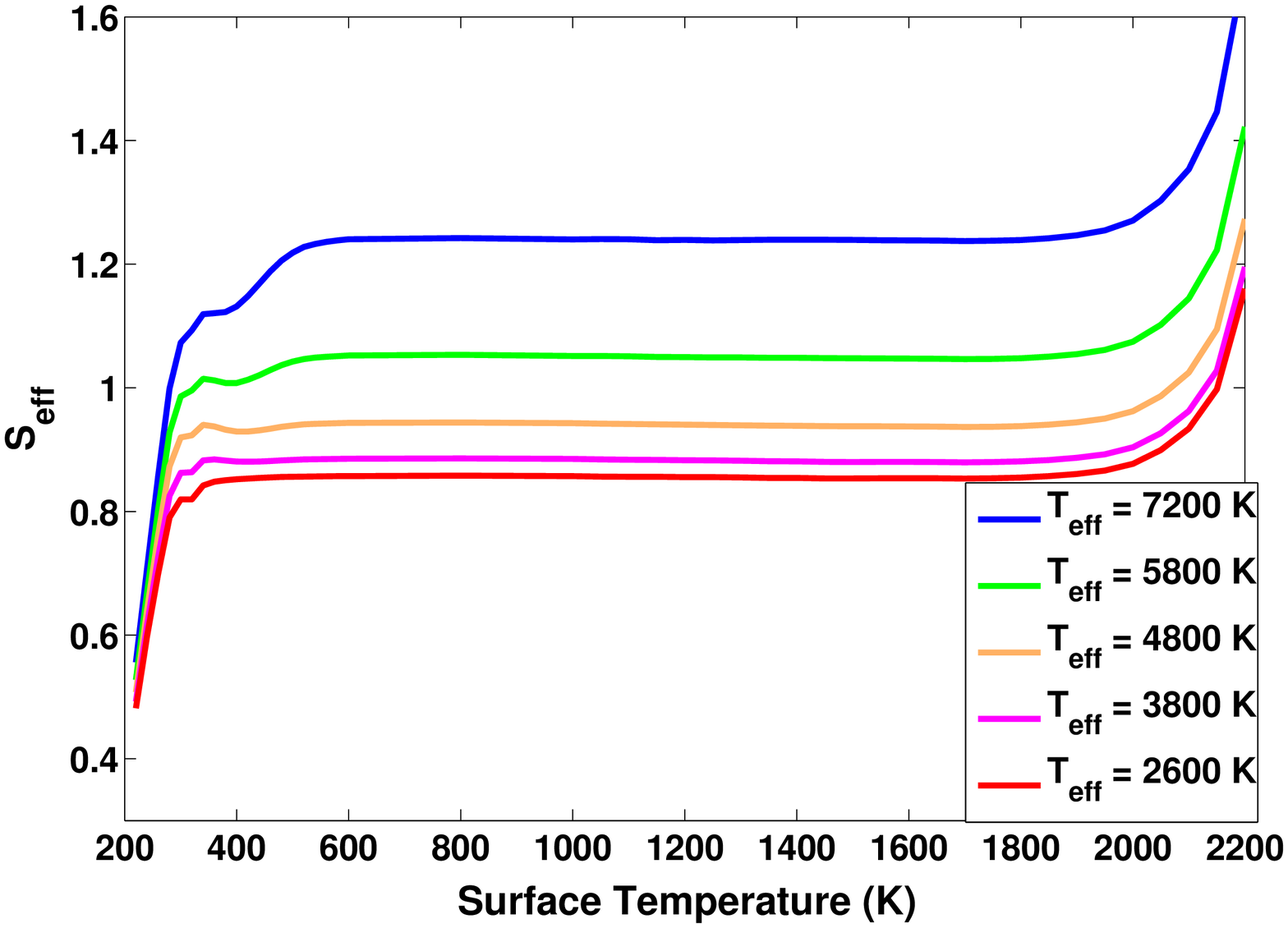}
}
\subfigure[] {
\label{FGMalbpco2}
\includegraphics[width=.50\textwidth]{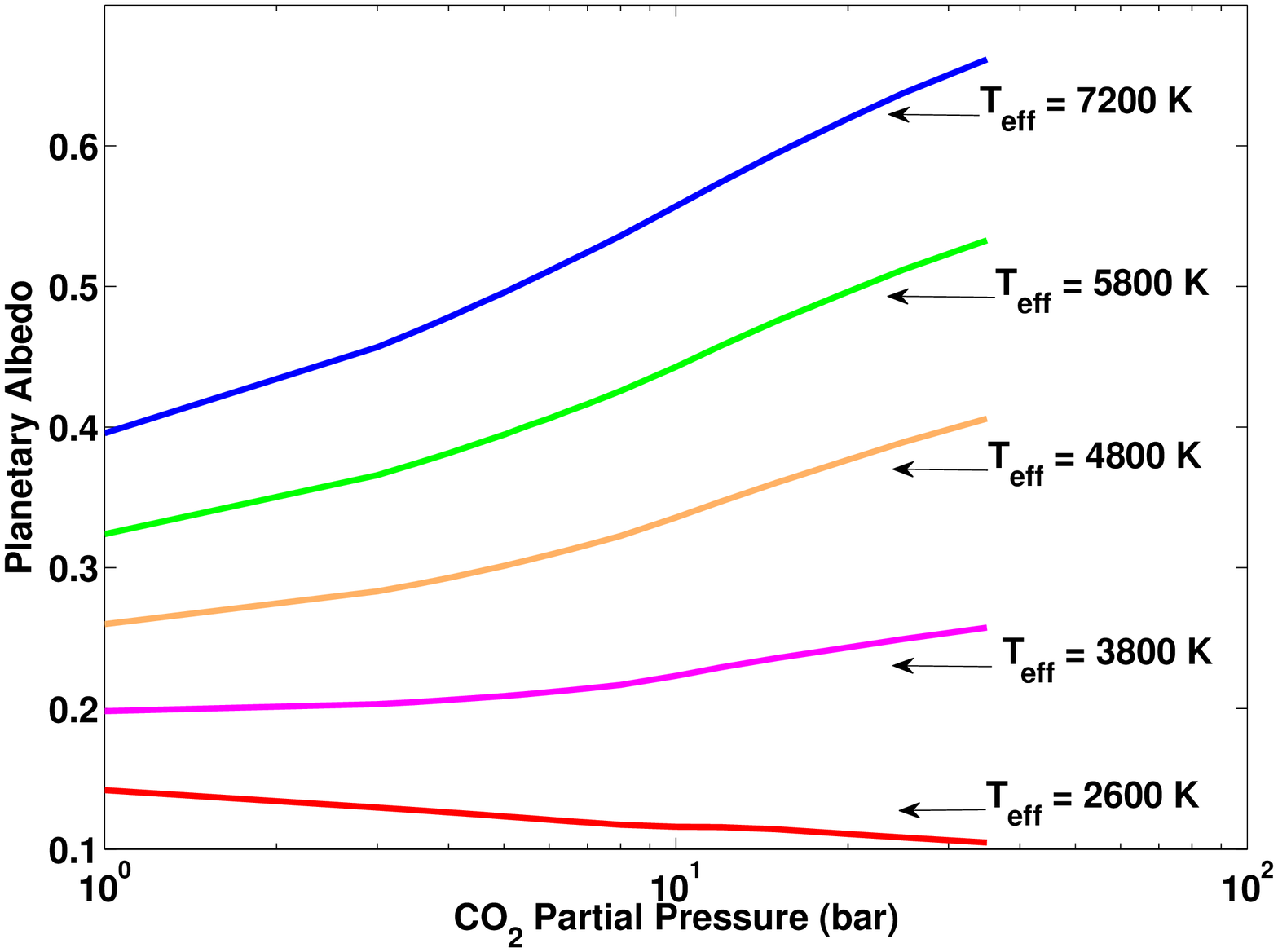}
}
\subfigure[] {
\label{seffpco2}
\includegraphics[width=.50\textwidth]{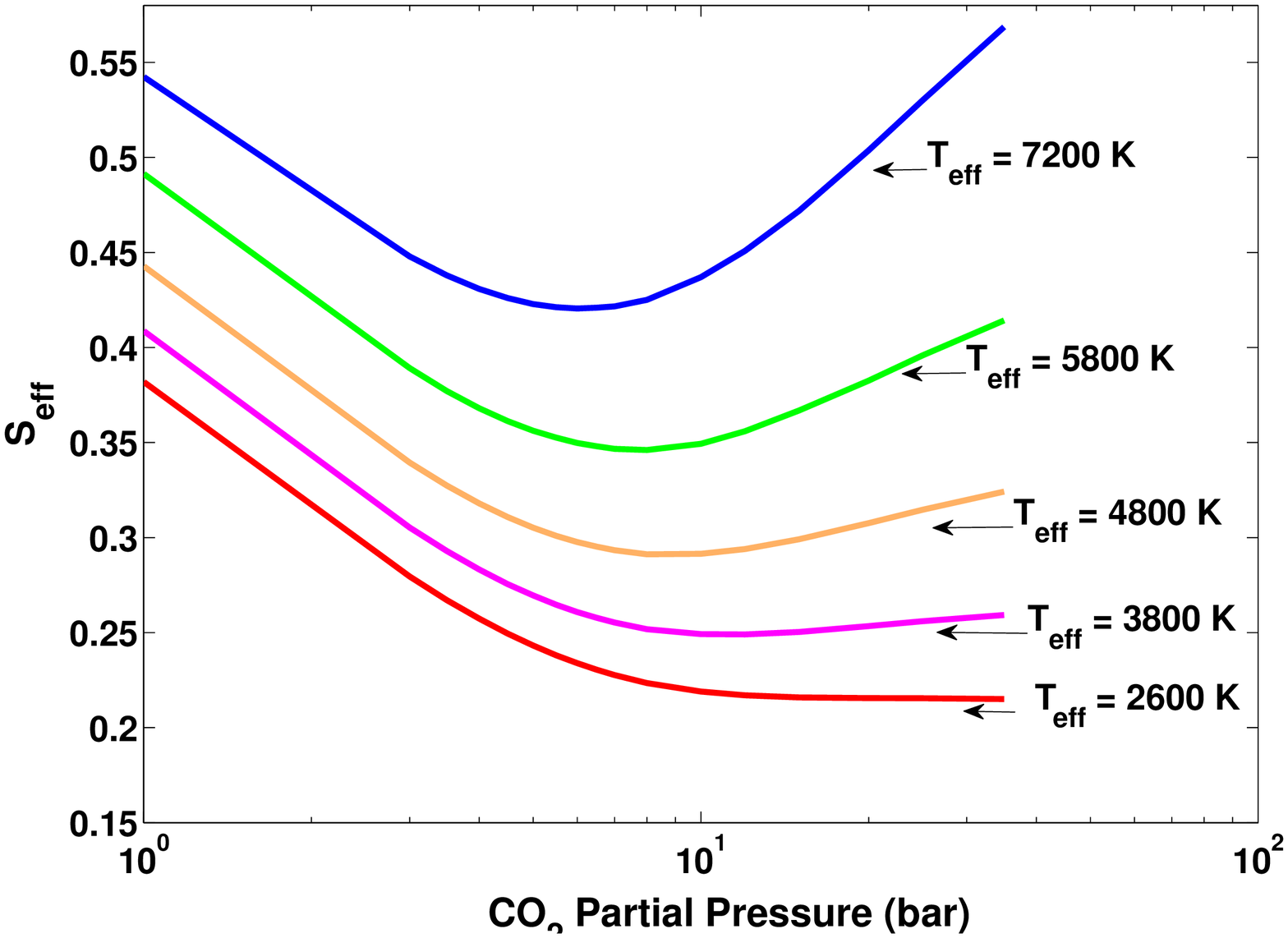}
}

\caption{Habitable zone calculations from our climate model for stellar effective
temperatures corresponding to F ($T_{eff} = 7200$ K), G (Sun), K ($T_{eff} = 4800$ K) 
and M ($T_{eff} = 3800$ K and 2600 K) spectral types.
The inner edge results are shown in the top row (Figs. \ref{albtg0} \& \ref{sefftg0}) and the
outer edge results are shown in the bottom row (Figs. \ref{FGMalbpco2} \& \ref{seffpco2}).}
\label{fgm}
\end{figure}

In Fig. \ref{fgm} we compare the results of our inner and outer edge HZ model calculations for
Sun to stars of different spectral types. Unless otherwise specified, we use the HITEMP 2010 
database for our inner edge calculations.
 The planetary albedo, shown in Fig. \ref{albtg0} (inner edge) and 
Fig. \ref{FGMalbpco2} (outer edge),  of an Earth-like planet is higher if the host star is
an F-star and lower if its primary is an M-star. The reason is that 
 the Rayleigh 
scattering cross section (which is proportional to $1/\lambda^{4}$) 
is on average higher for a planet around an F-star, as the star's Wien peak is bluer compared to the Sun. 
Second,
 $\h2o$ and $\co2$ have stronger absorption coefficients in the near-infrared than in the 
visible, so the amount of starlight absorbed by the planet's atmosphere increases as the
radiation is redder (as is the case for an M-star). Both effects are more pronounced
when the atmosphere is dense and full of gaseous absorbers. For a late M-star 
($T_{eff} = 2600$ K) most of its radiation is peaked around $1$ micron. Therefore,
the minimal amount of Rayleigh scattering and the high near-IR absorption by the planet's atmosphere combine to generate extremely low planetary albedos.

\thispagestyle{empty}

The changes in predicted planetary albedo can be translated into critical solar fluxes, as 
shown in Figs. \ref{sefftg0} (inner edge) and \ref{seffpco2} (outer edge). As discussed in
\S\ref{ohz}, $S_{eff}$ goes through a minimum near the OHZ because the atmosphere becomes optically thick
at all infrared wavelengths and, at the same time, the Rayleigh scattering due to 
$\co2$ condensation increases planetary albedo. Note that for a late M-star ($T_{eff} = 2600$ K)
Rayleigh scattering never becomes an important factor, and hence $S_{eff}$ asymptotically
reaches a constant value. 
The parameter $S_{eff}$  is directly calculated from our climate
model and is dependent on the type of star considered. Therefore, we have derived 
relationships between  HZ  stellar fluxes ($S_{eff}$) reaching the top of the 
atmosphere of an Earth-like planet and stellar effective temperatures ($T_{eff}$) applicable 
in the range $2600$ K $ \le T_{eff} \le 7200$ K. 
\begin{eqnarray}
\label{hzeq}
S_{eff} &=&  S_{eff\odot} + aT_{\star} + bT_{\star}^2 + cT_{\star}^3
                + dT_{\star}^4 
\end{eqnarray}
where $T_{\star} = T_{eff} - 5780$ K and the coefficients are listed in Table \ref{table4} for
various habitability limits\footnote{These coefficients can be downloaded in a machine readable
format from the electronic version of the journal. A fortran code is also available to calculate
 HZ stellar fluxes.}. 
The corresponding habitable zone distances can be calculated using
the relation:
\begin{eqnarray}
d &=& \biggl(\frac{L/L_{\odot}}{S_{eff}}\biggr)^{0.5} \mathrm{AU}
\label{dhz}
\end{eqnarray}
where $L/L_{\odot}$ is the luminosity of the star compared to the Sun. 

\begin{figure}[!hbp|t]
\subfigure[] {
\label{selsisflux}
\includegraphics[width=.85\textwidth]{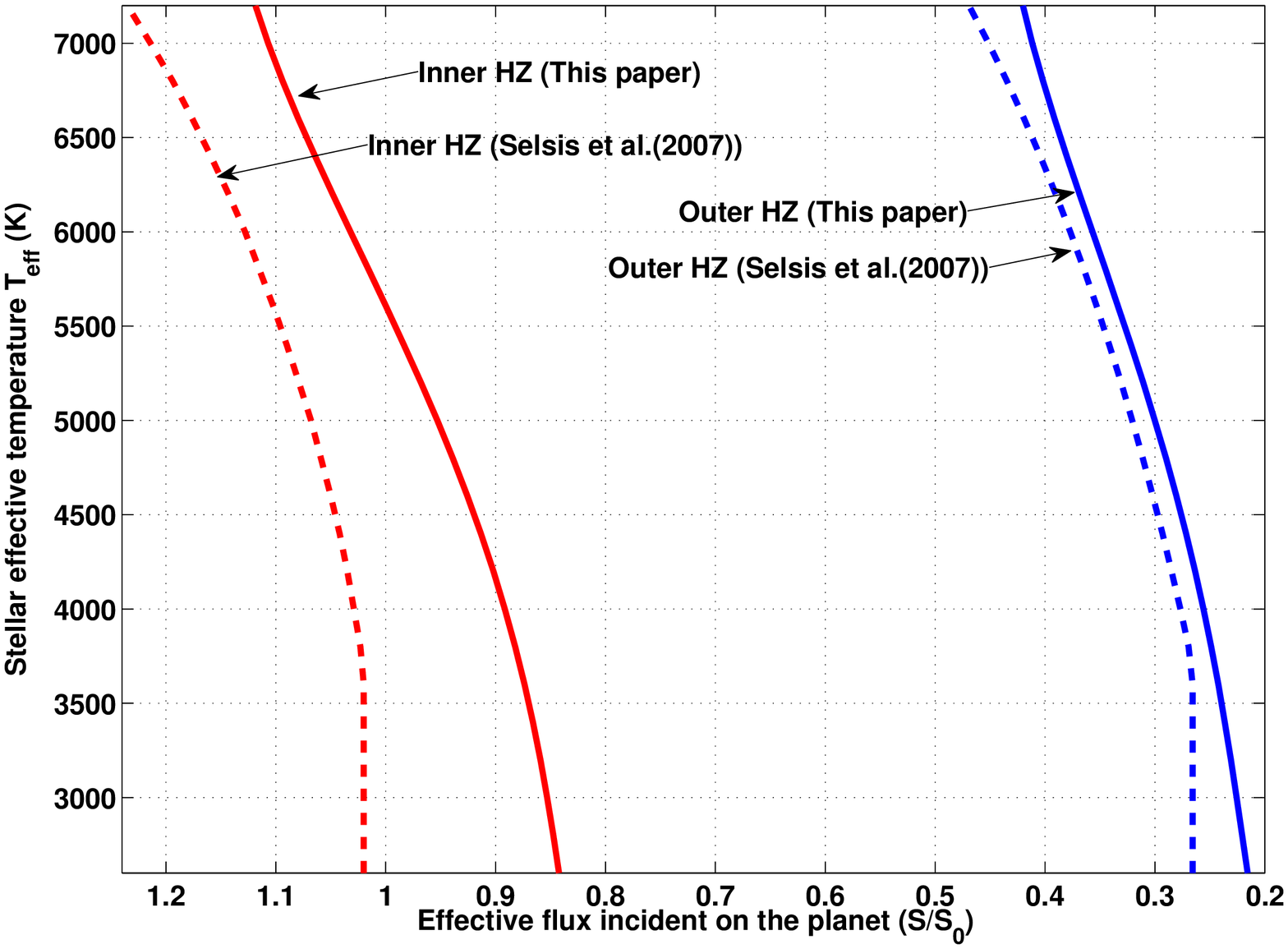}
}
\subfigure[] {
\label{selsisdis}
\includegraphics[width=.85\textwidth]{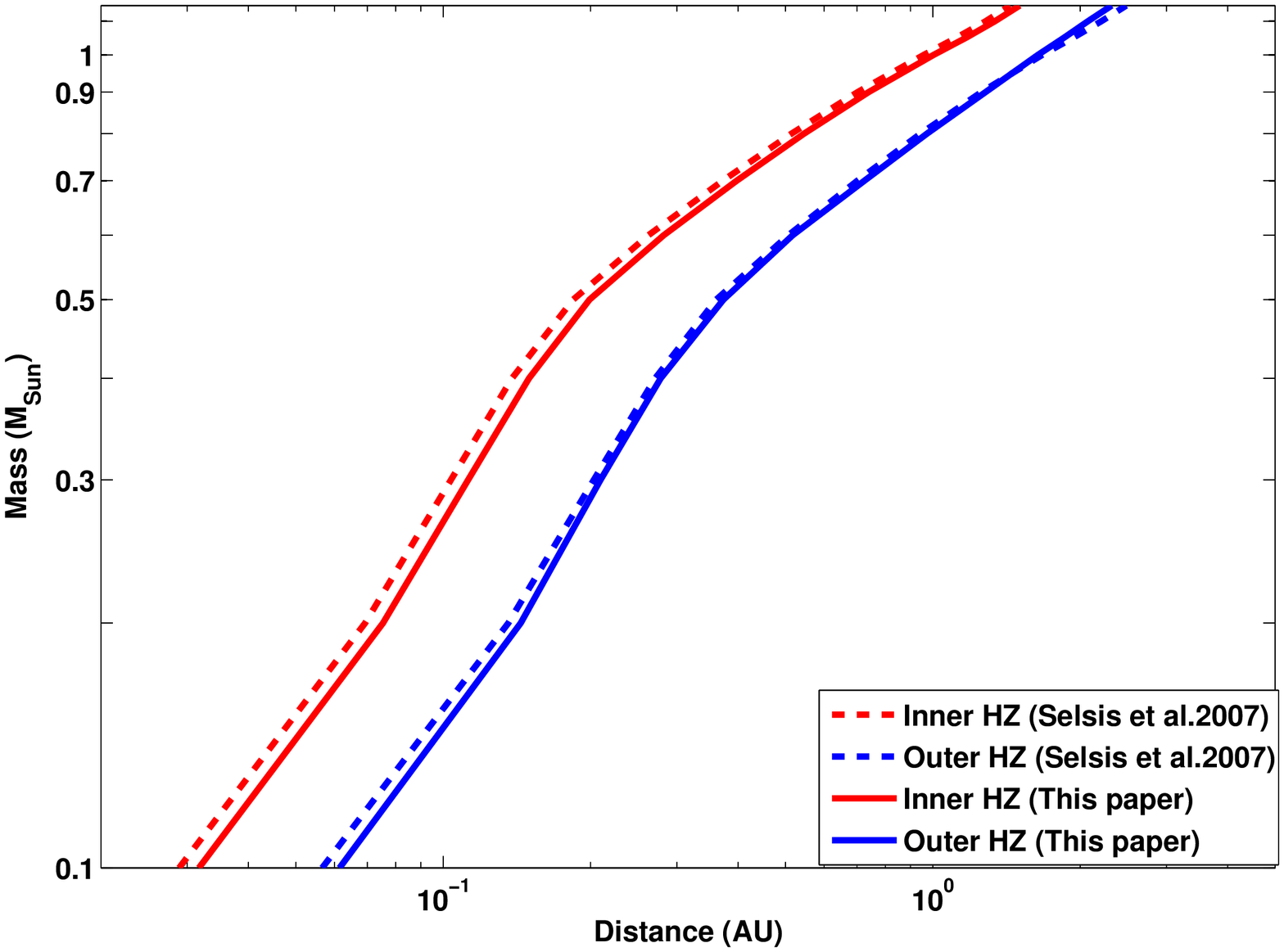}
}

\caption{Habitable zone fluxes (panel \ref{selsisflux}) and corresponding distances (panel \ref{selsisdis})
from our model (solid lines) compared to \cite{Selsis2007b} results (dashed lines) for different stellar
effective temperatures. The inner HZ fluxes 
from our model (red solid lines) are for the moist greenhouse case and outer edge fluxes 
(blue solid lines) are for the maximum greenhouse limit,
which are compared with $0 \%$ cloud cover limit from Table 2 of \cite{Selsis2007b}. The corresponding
continuous HZ distances for 5 Gyr are shown in panel \ref{selsisdis}.}


\label{selsiscompare}
\end{figure}

In Fig. \ref{selsiscompare} we compare HZ fluxes (and distances) calculated using Eqs.(\ref{hzeq}) $\&$
(\ref{dhz}) for the moist greenhouse case, with \cite{Selsis2007b} $0\%$ cloud results for different
stellar effective temperatures. As shown in Fig. \ref{selsisflux}, for low $T_{eff}$, 
there are large differences at the inner edge (dashed and solid red curves) between the models.
 This is because the spectrum of low-mass stars shifts towards the longer wavelengths, resulting in
more near-IR flux compared to high-mass stars. In both the models the atmosphere of a planet 
in the inner HZ is 
$\h2o$-dominated, and so there is strong absorption in the near-IR. Since our model uses the most recent
HITEMP database which has more $\h2o$ lines in the near-IR, 
the moist greenhouse limit occurs at a lower flux (farther from the star). Also, \cite{Selsis2007b}
assumed $T_{eff} =3700$ K for stars with temperatures below this value. This amplifies the differences,
as these low mass stars have their peak fluxes in near-IR.  These differences in inner habitable zone boundaries may become important
for present and upcoming planet finding surveys around M-dwarfs such as MEARTH \citep{Nutzman2008}
and Penn State's HPF \citep{Suvrath2012}, whose goal is to discover
potentially habitable planets around M-dwarfs.

The luminosity of a main sequence star evolves over time, and consequently the HZ 
distances (Eq.(\ref{dhz})) also 
change with time. One can calculate ``continuous'' HZ (CHZ) boundaries within which a planet remains
habitable for a specified length of time (we chose 5 Gyr). 
In Fig. \ref{selsisdis}, we show CHZ boundaries as a function of stellar mass for both our model and
\cite{Selsis2007b} model, taking into account the
stellar evolutionary models of \cite{Baraffe1998} for solar metallicity stars.  Noticeable 
differences between the two models are seen for low
mass stars near the inner edge (as also seen in Fig. \ref{selsisflux}. 
 The large differences in $S_{eff}$ from Fig. \ref{selsisflux} do not appear as pronounced in 
Fig. \ref{selsisdis}  because it is a log scale and also because
the CHZ distance is 
inversely proportional to the square root of  $S_{eff}$ (Eq.(\ref{dhz})).

In order to assess the potential habitability of recently discovered exoplanets, equilibrium
temperature ($T_{eq}$) has been used as a metric \citep{Borucki2011, Batalha2012}. 
Assuming an emissivity of $0.9$,
the ranges of HZ boundaries are taken to be
$185$ K $ \le T_{eq} \le 303$ K \citep{Kasting2011b}. We would like to stress that
the stellar fluxes ($S_{eff}$) provide a better metric for habitability 
than does $T_{eq}$. This is because $T_{eq}$ involves an assumption about $A_{B}$ 
($0.3$, usually) that is generally not valid. This value of $A_{B}$ is good for present
Earth around our Sun. For a planet around a late M-star, $A_{B}$ can vary from 0.01 near the
inner edge to 0.1 at the outer edge (see Fig. \ref{fgm}), depending on its location. 
Similarly, $A_{B}$ for an F-star
can range in between $0.38 - 0.51$ for the inner and outer edge, respectively. This changes the
corresponding $T_{eq}$,  and so a uniform criterion for HZ boundaries based on $T_{eq}$
cannot be determined. 


\begin{threeparttable}[h!]
\caption{Coefficients to be used in Eq.(\ref{hzeq}) to calculate habitable stellar fluxes, 
and corresponding habitable zones (Eq.(\ref{dhz})), for stars with 
$2600 \le T_{eff} \le 7200$ K. An ASCII file containing these coefficients can be downloaded 
 in the electronic version of the paper.}
\vspace{0.1 in}
\centering
\begin{tabular}{|c|c|c|c|c|c|}
\hline
Constant& Recent & Runaway & Moist &Maximum & Early\\
&Venus&Greenhouse&Greenhouse&Greenhouse&Mars\\
\hline
$S_{eff\odot}$ & 1.7753& 1.0512 &1.0140 &0.3438 & 0.3179 \\
&&&&&\\
$a$ & $1.4316 \times 10^{-4}$ & $1.3242 \times 10^{-4}$  &$8.1774 \times 10^{-5}$ &$5.8942 \times 10^{-5}$ & $5.4513 \times 10^{-5}$ \\
&&&&&\\
$b$ & $2.9875 \times 10^{-9}$& $1.5418 \times 10^{-8}$ &$1.7063 \times 10^{-9}$ &$1.6558 \times 10^{-9}$ & $1.5313 \times 10^{-9}$ \\
&&&&&\\
$c$ & $-7.5702 \times 10^{-12}$ & $-7.9895 \times 10^{-12}$ & $-4.3241 \times 10^{-12}$& $-3.0045 \times 10^{-12}$& $-2.7786 \times 10^{-12}$ \\
&&&&&\\
$d$ & $-1.1635 \times 10^{-15}$&$-1.8328 \times 10^{-15}$& $-6.6462 \times 10^{-16}$ & $-5.2983 \times 10^{-16}$& $-4.8997 \times 10^{-16}$ \\
&&&&&\\
\hline
\end{tabular}
\label{table4}
\end{threeparttable}

\thispagestyle{empty}

\section{Discussion}
\label{discussion}

\begin{figure}[!hbp|t]
\includegraphics[width=.95\textwidth]{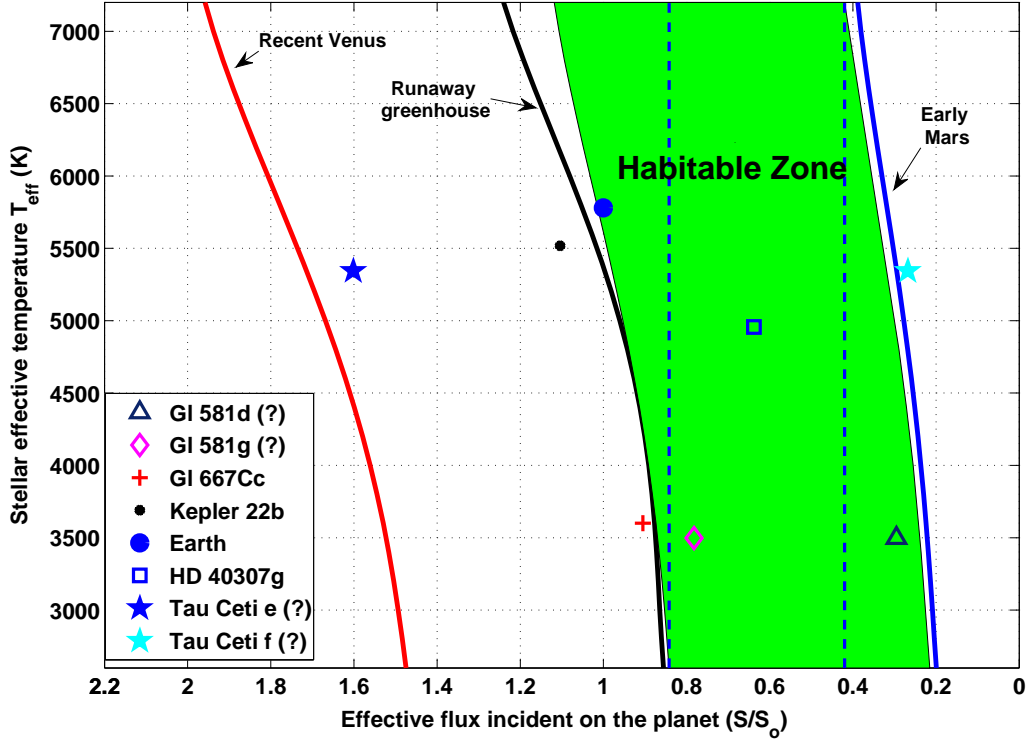}

\caption{Various cloud-free habitable zone (flux) boundaries for stars with different 
$T_{eff}$. 
The boundaries
of the green-shaded region are determined by the moist-greenhouse 
(inner edge, higher flux values) \& maximum greenhouse
(outer edge, lower flux values). A planet that receives stellar flux bounded by the two dashed vertical lines 
is in the HZ irrespective of the stellar type.
Some of the currently known exoplanets that are thought to be in the HZ by
previous studies are also shown. The `?' for Gl 581 and Tau Ceti system of planets 
imply that there is an ongoing discussion 
about their existence. For stars with T$_{eff} \lesssim 5000$ K, there is no clear distinction between
 runaway and moist-greenhouse limit.}

\label{observed}
\end{figure}

A straightforward application of the calculations presented in the previous sections is to apply them to
 currently known exoplanetary systems. Fig. \ref{observed} shows various habitable zone boundaries (expressed 
in terms of effective stellar flux) as a function of stellar $T_{eff}$.
The orbital parameters of the planets and stellar characteristics were obtained from
exoplanets.org \citep{Wright2011}.
 The green-shaded habitable region is bounded 
by the moist greenhouse limit (inner edge) and the maximum greenhouse limit 
(outer edge). Several
currently known terrestrial mass exoplanets that have been proposed to be in the HZ by various 
studies are also shown. 
An important insight that can be obtained from this figure (and one that cannot be seen in
the HZ distance plot, Fig. \ref{selsisdis}) is that a terrestrial mass planet that lies 
within the two vertical dashed-lines in the green
shaded region is in the HZ irrespective of the type of star it is orbiting. The
corresponding flux boundaries for which a rocky planet is `definitely' in the HZ  are
$0.842$ and $0.42$. Currently, two exoplanets fall within this 
region, HD 40307g \citep{Tuomi2012a} and Gl 581g \citep{Vogt2010,Vogt2012}.
The detection of planets and orbital parameters for Gl581 is complicated by the low amplitudes of 
the signal, stellar activity, and possible red noise. We have included a `?' for Gl 581 system 
of planets in the plot, indicating that there is an active ongoing discussion in the literature 
about the number of planets in this system and their exact orbital parameters  
\citep{Vogt2010, Pepe2011b, Vogt2012, Baluev2012}.
Furthermore, for stars with T$_{eff} \lesssim 5000$ K, there is no clear distinction between
runaway greenhouse and the moist-greenhouse limits. The reason is that for these stars, there are
more photons available in the IR part of the spectrum, where $\h2o$ is a good absorber. Therefore
a planet with $\h2o$ dominated atmosphere quickly goes into runaway  once it reaches 
the moist-greenhouse limit. 
Note that another suggested HZ candidate planet, HD 85512b 
\citep{Pepe2011a}, receives more than 5 times the stellar flux received by our Earth,
placing it even beyond the most liberal (`recent Venus') estimate of the inner edge.
Hence, it is very likely that this planet is not in the HZ of its star.

A question of importance to the exoplanet community is which HZ limits to choose when identifying
potentially habitable planets.
For current RV surveys and  {\it Kepler} mission one should use the most conservative limits
(moist greenhouse and maximum greenhouse), because this will give a lower limit on $\eta_{\oplus}$,
the fraction of sun-like stars that have at least one planet in the habitable zone \citep{exoptf2008}.
If one is interested in designing a future flagship mission, such as 
{\it Terrestrial Planet Finder} (TPF) or {\it Darwin}, then using these conservative limits 
(which results in a lower limit on $\eta_{\oplus}$) ensures that the telescope is not undersized. 
If, however, one was analyzing data obtained from such a telescope, the most optimistic 
limits (recent Venus and early Mars) should be used because 
one would not want to miss out on any potentially habitable planets.

\begin{figure}[!hbp|t]
\includegraphics[width=.95\textwidth]{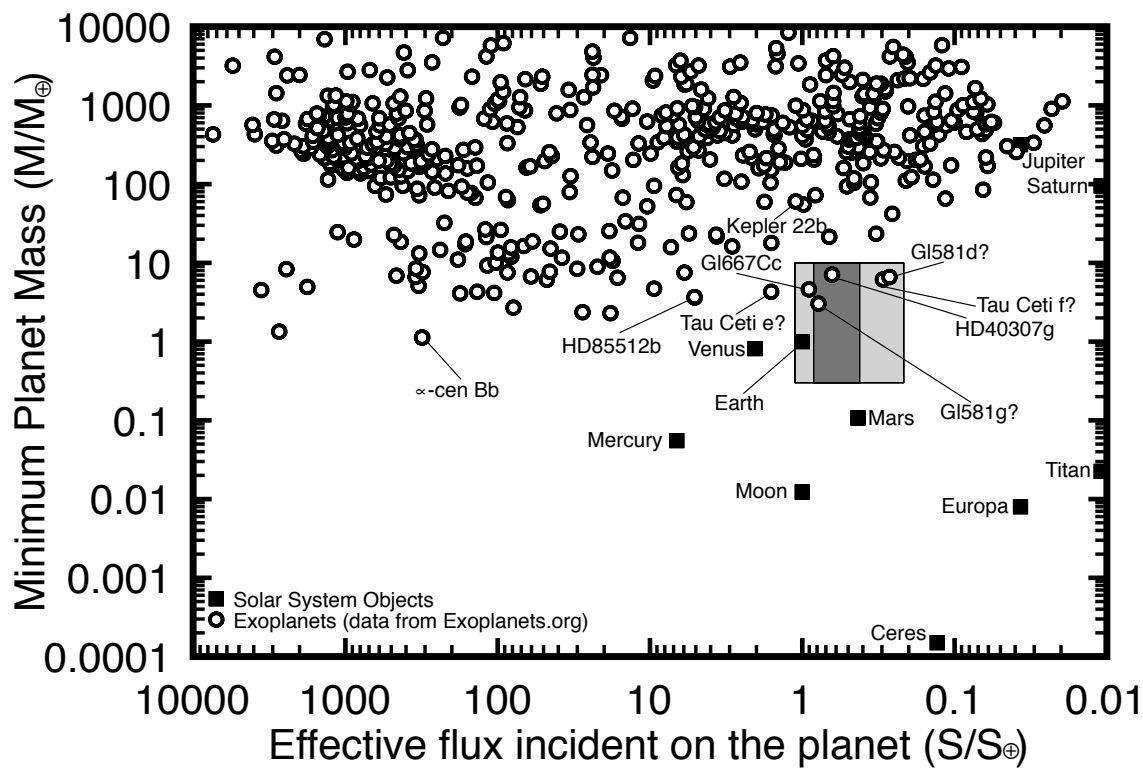}

\caption{HZ boundaries combining observable stellar and planetary parameters. Terrestrial
planets ($0.3$M$_{\oplus}$ - $10$M$_{\oplus}$, y-axis limits)
within the dark shaded region are in the HZ irrespective of the stellar type. For those
planets in the light grey region, one needs to know the stellar type to determine if they are
in the HZ. For example, the Earth would not be habitable if it received its current
incoming amount of energy from a cooler-type star, but it is (obviously)
habitable in its current orbit around a G-type star. For planets such as
this, the star's energy distribution matters when considering
habitability.}

\label{exoplanetsplot}
\end{figure}

In Fig. \ref{exoplanetsplot}, we show  the incident stellar flux as a function of planetary mass
for the currently known exoplanets. The masses are obtained from exoplanets.org when
available.
Also shown are
habitable zone flux boundaries calculated from Eq.(\ref{hzeq}) for terrestrial mass 
planets ($1$M$_{\oplus}$ - $10$M$_{\oplus}$).
For the outer box (light grey), the upper bound on the flux 
is taken to be the moist greenhouse limit for a star with $T_{eff} = 7200$ K and
the lower bound is the maximum greenhouse limit for a star with $T_{eff} = 2600$ K.
These are indicated by the diagonally opposite points on the green shaded region in 
Fig. \ref{observed}. For the inner box (dark grey), the flux limits are the dashed lines in
the green shaded region of Fig. \ref{observed}. The significance of this plot is that 
terrestrial planets in the inner box must be in the HZ, irrespective of the 
stellar spectral type. Mars, if it were more massive,
would be in the HZ around any main sequence star with $2600 \le T_{eff} \le 7200$ K.
For planets that are outside the dark grey region, but inside the 
light grey (for example, Earth) one needs to know the host star's spectral type (or 
$T_{eff}$) to determine if that planet is in the HZ. Fig. \ref{exoplanetsplot} combines
observable stellar and planetary parameters to further constrain HZ boundaries for
extrasolar planets.

 Many of the currently known exoplanets have non-zero eccentricities, which can 
carry some of them (and their possible moons) in and out of the HZ. The incident
stellar flux on these eccentric planets has extreme variations between periastron and
apoastron ($[(1+e)/(1-e)]^{2}$). \cite{WP2002} show that, provided that an ocean is present to act as a heat capacitor, it is primarily the time-averaged flux 
$<S_{eff}^{~\prime}> $ that affects the habitability over an eccentric 
orbit \citep{Kopparapu2009, Kopparapu2010}. Mathematically:
\begin{eqnarray} 
<S_{eff}^{~\prime}> &=& \frac{S_{eff}}{(1 - e^2)^{1/2}}
\label{ecchz}
\end{eqnarray} 
Here, $S_{eff}$ is the effective flux from circular orbit (Eq.(\ref{hzeq})).
Planets with high orbital eccentricities ($e \sim \ge 0.1$) have higher average orbital 
flux. This may help eccentric planets near the outer edge of the HZ maintain
habitable conditions. However, obliquity variations can influence the geographical distribution
of irradiation \citep{Spiegel2008, Spiegel2009, Dressing2010} and may change habitable conditions.

 Earth itself appears to be perilously 
close to the moist greenhouse limit
($S_{eff} = 1.015$, blue filled circle in Fig. \ref{observed}). However, this apparent instability is deceptive, because the calculations
do not take into account the likely increase in Earth's albedo that would be caused by water clouds on a warmer Earth.
Furthermore, these calculations assume a fully saturated troposphere that maximizes the greenhouse
effect. For both reasons, it is likely that the actual HZ inner edge is closer to the Sun than our moist
greenhouse limit indicates. Note that the moist greenhouse in our model occurs at a surface 
temperature of 340 K. The current average surface temperature of the Earth is only 288 K. 
Even a modest (5-10 degree) increase in the current surface temperature could have devastating affects
on the habitability of Earth from a human standpoint. 
 Consequently, though we identify the moist greenhouse limit
as the inner edge of the habitable zone, habitable conditions for humans could disappear well before Earth reaches this limit.

Additional uncertainty about habitability of planets around late-K and M stars.
($T_{eff}  <= 4000$ K) comes from the fact that planets within the HZs of these stars are expected to be tidally
locked \citep{Dole1964, Peale1977,Kasting1993, Dobrovolskis2009}. If the planet's 
orbital eccentricity is small, this can result in synchronous rotation,
in which one side of a planet always faces the star (as the Moon does to the Earth).
Climates of synchronously rotating planets are not well approximated by 1-D, globally
averaged models. Previous work has shown that such planets may indeed be habitable
\citep{Joshi1997, Joshi2003, Edson2011};
however, systematic exploration of
synchronously rotating planets in different parts of the HZ has not been attempted. Even
before doing these calculations, we can predict that planets near the outer edge of the HZ,
with their expected dense $\co2$ atmospheres, should be more effective at transporting heat
around to their night sides, and hence should have a better chance of being habitable.

Given that survey like HPF and CARMENES will specifically target mid-late M dwarfs, our future 
work will include estimating the HZ boundaries of individual targets in detail. Ongoing work by 
our team \citep{Terrien2012} is yielding low resolution NIR spectra from the infrared Telelescope Facility 
(IRTF) to be used to derive stellar metallicities as well as yielding more realistic flux 
distributions and temperatures for use in the modeling. We anticipate having this information 
for $\sim$ 650 M dwarfs drawn from the J$<10$ \cite{LG2011}catalog, and have
applied for time to observe $\sim$ 300 more. 
Estimates of luminosities will be derived using photometric and spectroscopic distances 
for now (in cases where parallax measurements are absent), but eventually GAIA 
\citep{Perryman2001} will 
yield very precise parallaxes (and by extension precise luminosities) for all these target stars.


Recent discoveries by both the {\it Kepler} mission and RV surveys
 have shown that planets can exist in
stable orbits around multiple star systems \citep{Doyle2011, Welsh2012, Orosz2012,
Dumusque2012}. The HZs of these stars could potentially host terrestrial planets,
which are at the threshold of current detection techniques. Indeed, the discovery
 of Kepler 47c \citep{Orosz2012} which is 4.6 times the size of the Earth's radii 
 in the HZ is a step closer to discovering  rocky planets in the HZ of multiple star
systems. 
\cite{Dumusque2012} have recently published a possible detection of a 1.1 Earth mass (minimum) 
planet in a 3.236 day orbit around $\alpha$  Centauri system.
So, this system should now be a prime 
target for further observations to discover habitable planets. 
Formation of dynamically stable terrestrial planets in the HZs of 
multiple star systems has been
studied before \citep{Whitmire1998, Holman1999, Haghighipour2007} and several
studies estimated HZ boundaries around these types of systems \citep{Eggl2012a, 
Eggl2012b, KH2012} using \cite{Kasting1993} model.
Our updated model results
from Fig. \ref{observed} or  Eqs.(\ref{hzeq}) and (\ref{dhz}) 
could change these estimates significantly.

Our new model results could also directly affect estimates of $\eta_{\earth}$.
Recent analysis of {\it Kepler} data \citep{Traub2012} and RV surveys \citep{Bonfils2011} concluded 
that $\eta_{\earth} \sim 0.34 - 0.4$. These values were based either on the \cite{Kasting1993} model
\citep{Traub2012} or the \cite{Selsis2007b} results \citep{Bonfils2011}.
Our new HZ limits could impact these estimates significantly. In particular, 
 there are large differences between \cite{Selsis2007b} calculations and our model results 
for low mass stars. The estimate of  $\eta_{\earth}$ by \cite{Bonfils2011} is obtained by using 
\cite{Selsis2007b} relationships for planets orbiting M-stars. 
Thus, this value may need to be re-evaluated.


\thispagestyle{empty}

\thispagestyle{empty}

\section{Conclusions}
\label{conclusions}
We have obtained new estimates for HZs around F, G, K and M main-sequence stars 
by (1) updating $\h2o$ and $\co2$ absorption coefficients in the \cite{Kasting1993} 
1-D radiative-convective cloud-free climate model with the most 
recent LBL databases: HITRAN 2008 and HITEMP 2010, (2) correcting the $\h2o$ Rayleigh scattering
coefficient, and (3) updating $\co2$ collision-induced absorption
coefficients. These changes affect the inner and outer edges of the HZ, respectively.

Our revised model predicts that the moist greenhouse limit for our Sun, which defines
the inner edge of the HZ, is at 0.99 AU. The outer edge of the HZ, where gaseous $\co2$ 
produces its maximum greenhouse effect, is at 1.70 AU. Although it appears that Earth
is perilously close to the inner HZ edge, in reality, cloud feedback and low upper tropospheric relative humidity act to stabilize Earth’s climate. Theoretical studies and observational surveys that depend on these limits should use the updated values. We have also estimated HZ boundaries for M stars with $T_{eff}$ as
low as $2600$ K, which are primary targets for ongoing surveys such as 
{\it Habitable Zone Planet Finder}   and {\it MEARTH} to discover potential habitable
planets.

 We also showed that the effective stellar flux provides a better 
criterion in determining the HZ limits, than equilibrium temperature.
Accordingly we have derived
a generalized expression to calculate these fluxes for stars of different spectral types.
Our results show that some of the extrasolar planets that were previously thought to be
within the HZ may not be in that region. Thus, our HZ estimates can be used to  
narrow the target list for eventual characterization missions, such as {\it JWST}, to
identify potential biomarkers on habitable planets.

Author Contribution: R.K \& R.R contributed equally to this work.
An interactive webpage to obtain HZs and a FORTRAN code is available here:
\url {http://depts.washington.edu/naivpl/content/hz-calculator}. The FORTRAN code is also
available in the electronic version of the paper.

\acknowledgements

The authors are especially grateful to David Crisp for his invaluable comments, suggestions, 
and for answering our radiative transfer questions during the preparation of this work.
The authors would like to thank the referee Robin Wordsworth for his constructive 
comments which improved the manuscript. Note added in proof: R. Wordsworth points out that his 
model uses increased vertical resolution in the lower atmosphere and that his results were found 
to be insensitive to further increases in resolution. All such calculations should be tested 
to determine whether they are robust to this issue.
We also thank Colin Goldblatt, 
David Paynter, Richard Freedman, Itay Halevy, Eli Mlawer and
Martin Cohen (U.C Berkeley) for their helpful discussions. 
The authors acknowledge the Research Computing and Cyberinfrastructure
unit of Information Technology Services at The Pennsylvania State
University for providing advanced computing resources and services that
have contributed to the research results reported in this paper. {\url {
http://rcc.its.psu.edu}}.  This work was also facilitated through the use of 
advanced computational, storage, and networking infrastructure provided by the 
Hyak supercomputer system, supported in part by the University of Washington eScience Institute.
This research has made use of the Exoplanet Orbit Database
and the Exoplanet Data Explorer at exoplanets.org.

R. K, R. R, J.F.K and SDDG gratefully acknowledge funding from NASA Astrobiology
 Institute's  Virtual 
Planetary Laboratory lead team, supported by NASA under cooperative agreement
NNH05ZDA001C, and the Penn State Astrobiology Research Center.
V.E. acknowledges the support of the ITAAC project (Impact du Trafic A\'{e}rien 
sur l'Atmosph\'{e}re et le Climat), funded by the Fondation Sciences et Technologies pour 
l'A\'{e}oronautique et l'Espace (STAE), Toulouse, France, within the R\'{e}seau Th\'{e}matique 
de Recherche Avanc\'{e}e(RTRA), and support from the European Research Council (Starting Grant 209622: E3ARTHs).
SM acknowledges support from NSF AST1006676, AST1126413, PSARC, and the NASA NAI. The Center for Exoplanets and Habitable Worlds is supported by the
Pennsylvania State University, the Eberly College of Science, and the
Pennsylvania Space Grant Consortium. SDDG was also supported by the Oak Ridge Associated
Universities NASA Postdoctoral Management Program, and did much of his
work on this project while in residence at NASA Headquarters.



\begin{thebibliography}{}

\bibitem[Abbot et al.(2012)]{Abbot2012}
Abbot,D. S., Cowan, N. B., \& Ciesla, F. J. 2012. \apj, 756, article id. 178

\bibitem[Abe et al.(2011)]{Abe2011}
Abe, Y., Abe-Ouchi, A., Sleep, N. H., \& Zahnle, K. J. 2011. Astrobiology, 11, 443

\bibitem[Allard et al.(2003)]{Allard2003}
Allard, F., Guillot, T., \& Ludwig, H. G. et al. 2003, {\it Brown Dwarfs, Proceedings of IAU Symposium},
211, 20-24 May, 2002, University of Hawaii, Honolulu, Hawai

\bibitem[Allard et al.(2007)]{Allard2007}
Allard, F., Allard, N. F., \& Homeier, D. et al. 2007, \aap, 474, L21


\bibitem[Allen (1976)]{Allen1976}
Allen, C. 1976, Astrophysical Quantities (University of London: The Athlone
Press)

\bibitem[Bahcall et al.(2001)]{Bahcall2001}
Bahcall, J. N., Pinsonneault, M. H., \& Basu, Sarbani. 2001. \apj, 555, 990

\bibitem[Baluev(2012)]{Baluev2012}
Baluev, R. V. 2012. submitted to MNRAS.

\bibitem[Baranov et al.(2004)]{Baranov2004}
Baranov. Y. I., Lafferty, W. J., \& Fraser, G. T. 2004. {\it Journal of Molecular Spectroscopy},
228, 432

\bibitem[Baraffe et al.(1998)]{Baraffe1998}
Baraffe, I., Chabrier, G., Allard, F., \& Hauschildt, P. 1998, \aap, 337, 403

\bibitem[Batalha et al.(2012)]{Batalha2012}
Batalha, N. M., Rowe, J. F., Bryson, S. T. et al. 2012. submitted to ApJS, arXiv:1202.5852

\bibitem[Bender et al.(2012)]{Bender2012}
Bender, C. F., Mahadevan, S., Deshpande, R. 2012. \apjl, 751, L31

\bibitem[Bezard et al.(1990)]{Bezard1990}
Bezard, B., Debergh, C., Crisp, D., \& Maillard, J. P. 1990, {\it Nature}, 345, 508

\bibitem[Bibring et al.(2006)]{Bibring2006}
Bibring, J.-P., Langevin, Y., Mustard, J. F., et al. 2006, {\it Science}, 312, 400


\bibitem[Bonfils et al.(2011)]{Bonfils2011}
Bonfils, X., Delfosse, X., Udry, S. et al. 2011. submitted \aap, arXiv:1111.5019

\bibitem[Borucki et al.(2011)]{Borucki2011}
Borucki, W. J., Koch, D. G., Basri, G. et al. 2011. \apj, 736, article id. 19 

\bibitem[Borucki et al.(2012)]{Borucki2012}
Borucki, W. J., Koch, D. G., Batalha, N. et al. 2012. \apj, 745, article id. 120


\bibitem[Bucholtz (1995)]{Bucholtz1995}
Bucholtz, A. 1995, {\it Appl. Opt.}, 34, 2765

\bibitem[Burch et al.(1969)]{Burch1969}
Burch, D. E., Gryvnak, D. A., Patty, R. R., \& Bartky, C. E. 1969. {\it Opt. Soc. Am.},
59, 267

\bibitem[Clampin et al.(2007)]{Clampin2007}
Clampin, M., Valenti, J., Deming, D. 2007. Detection of Planetary Transits with the James Webb Space Telescope. ExoPTF whitepaper. 

\bibitem[Clough et al.(1989)]{CKD1989}
Clough, S. A., Kneizys, F. X., and Davies, R. W. 1989. {\it Atmospheric Research}, 23, 229

\bibitem[Clough \& Iocono(1995)]{CI1995}
Clough, S. A., \& Iacono, M. J. 1995. {\it J. Geophys. Res}, 100, 16519

\bibitem[Colaprete \& Toon(2003)]{CT2003}
Colaprete, A., \& Toon, O. B. 2003. {\it Journal of Geophysical Research (Planets)}, 108, 6-1

\bibitem[Crisp(1997)]{Crisp1997}
Crisp, D. 1997. {\it Geophysical Research Letters}, 24, 571


\bibitem[Deming et al.(2009)]{Deming2009}
Deming, D., Seager, S., Winn, J. et al. 2009. PASP, 121, 952

\bibitem[Dobrovolskis (2009)]{Dobrovolskis2009}
Dobrovolskis, A. R. 2009. {\it Icarus}, 204, 1

\bibitem[Dole (1964)]{Dole1964}
Dole, S. H. 1964. {\it Habitable Planets for Man}. New York: Blaisdell Publishing. 158 pp

\bibitem[Doyle et al.(2011)]{Doyle2011}
Doyle, L. R., Carter, J. A., Fabrucky, D. C. et al. 2011. {\it Science}, 333, 1602

\bibitem[Dressing et al.(2010)]{Dressing2010}
Dressing, C. D., Spiegel, D. S., Scharf, C. A., Menou, K., \& Raymond, S. N. 2010. \apj, 721, 1295

\bibitem[Dumusque et al.(2012)]{Dumusque2012}
Dumusque, X., Francesco, P., Christophe, L. et al. 2012. {\it Nature}, doi:10.1038/nature11572


\bibitem[Edl\'{e}n(1996)]{Edlen1996}
Edl\'{e}n, B. 1966. {\it Metrologia}, 2, 71

\bibitem[Edson et al.(2003)]{Edson2011}
Edson, A., Lee, S., Bannon, P.,  Kasting, J. F., \& Pollard, D. 2011. {\it Icarus}, 212, 1

\bibitem[Eggl et al.(2012a)]{Eggl2012a}
Eggl, S., Pilat-Lohinger, E., Georgakarakos, N. et al. 2012. \apj, 752, article id.  74

\bibitem[Eggl et al.(2012b)]{Eggl2012b}
Eggl, S., Pilat-Lohinger, E., Funk, B. 2012. arXiv:1210.5411

\bibitem[Forget \& Pierrehumbert(1997)]{FP1997}
Forget, F., \& Pierrehumbert, R.T., 1997. {\it Science}, 278, 1273

\bibitem[Forget et al.(2012)]{Forget2012}
Forget, F., Wordsworth, R. W., Millour, E. et al. 2012. Icarus accepted. arXiv:1210.4216

\bibitem[Fukabori et al.(1986)]{Fukabori1986}
Fukabori, M., Nakazawa, T., \& Tanaka, M. 1986. {\it JQSRT}, 36, 265 

\bibitem[Goldblatt \& Zahnle(2011)]{GZ2011}
Goldblatt, C., \& Zahnle, K. 2011. {\it Nature}, 474, 7349

\bibitem[Gough(1981)]{Gough1981}
Gough, D. O. 1981. ESA and European Physical Society, ESLAB Symposium on Physics of Solar Variations, 14th, Scheveningen, Netherlands. Solar Physics, 74, 21

\bibitem[Gruszka \& Borysow(1994)]{GB1994}
Gruszka, M. \& Borysow, A. 1994. J. Chem. Phys, 101, 3573

\bibitem[Gruszka \& Borysow(1997)]{GB1997}
Gruszka, M. \& Borysow, A. 1997. {\it Icarus}, 129, 172

\bibitem[Haghighipour \& Raymond(2007)]{Haghighipour2007}
Haghighipour, N., \& Raymond, S. 2007, \apj, 666, 436

\bibitem[Halevy et al.(2009)]{Halevy2009}
Halevy, I., Pierrehumbert, R. T., \& Schrag, D. P. 2009. {\it Journal of Geophysical Research},
114, D18112


\bibitem[Haqq-Misra et al.(2008)]{Jacob2008}
Haqq-Misra, J. D., Domagal-Goldman, S. D., Kasting, P. J. \& Kasting, J. F. 2008. {\it Astrobiology}, 8, 1127

\bibitem[Hart(1978)]{Hart1978}
Hart, M. H. 1978, {\it Icarus}, 33, 23

\bibitem[Holman \& Wiegert(1999)]{Holman1999}
Holman, M .J., \& Wiegert, P. A. 1999. \aj, 117, 621

\bibitem[Huang(1959)]{Huang1959}
Huang, S. S. 1959, American Scientist, 47, 397


\bibitem[Joshi et al.(1997)]{Joshi1997}
Joshi, M. M., Haberle R. M., Reynolds R. T. 1997. {\it Icarus}, 129, 450

\bibitem[Joshi(2003)]{Joshi2003}
Joshi, M. M. 2003. {\it Astrobiology}, 3, 415

\bibitem[Kaltengger \& Traub(2009)]{KT2009}
Kaltenegger, L., \& Traub, W. 2009. \apj. 698, 519

\bibitem[Kaltenegger et al.(2011a)]{Kaltenegger2011a}
Kaltenegger, L., Segura, A., \& Mohanty, S. 2011a. \apj. 733. id. 35

\bibitem[Kaltenegger et al.(2011b)]{Kaltenegger2011b}
Kaltenegger, L., \& Sasselov, D. 2011b. \apjl, 736, L25

\bibitem[Kane \& Hinkel(2012)]{KH2012}
Kane, S. R, \& Hinkel, N. R. 2012. \apj accepted. arXiv:1211.2812

\bibitem[Kasting et al.(1984)]{Kasting1984}
Kasting, J. F., Pollack, J. B., Crisp, D. 1984. {\it J. Atmos. Chem.}, 1, 403

\bibitem[Kasting \& Ackerman(1986)]{KA1986}
Kasting, J. F., \& Ackerman, T. P. 1986. {\it Science}, 234, 1383

\bibitem[Kasting(1988)]{Kasting1988}
Kasting, J., F. 1988, {\it Icarus}, 74, 472

\bibitem[Kasting(1991)]{Kasting1991}
Kasting, J., F. 1991, {\it Icarus}, 94, 1

\bibitem[Kasting et al.(1993)]{Kasting1993}
Kasting, J., F., Whitmire, D., P., \& Reynolds. R. T. 1993, {\it Icarus}, 101, 108

\bibitem[Kasting(2011b)]{Kasting2011b}
Kasting, J. F. 2011b. Joint Meeting of the Exoplanet and Cosmic Origins Program
Analysis Groups (ExoPAG and COPAG), April 26, 2011, Baltimore, MD,
\url {http://exep.jpl.nasa.gov/exopag/exopagCopagJointMeeting/}

\bibitem[Kato et al.(1999)]{Kato1999}
Kato, S., Ackerman, T. P., Mather, J. H. et al. 1999. {\it J. Quant.
Spectrosc. Radiat. Transf.}, 62, 109

\bibitem[Kitzmann et al.(2011a)]{Kitzmann2011a}
Kitzmann, D., Patzer, A. B. C., Von Paris, P. P., Godolt, M., \& Rauer, M. 2011a. \aap. 531, id.A62

\bibitem[Kitzmann et al.(2011b)]{Kitzmann2011b}
Kitzmann, D., Patzer, A. B. C., Von Paris, P. P., Godolt, M., \& Rauer, M. 2011b. \aap. 534, id.A63


\bibitem[Kopparapu et al.(2009)]{Kopparapu2009}
Kopparapu, R., Raymond, S. N., Barnes, R. 2009. \apjl, 695, L181

\bibitem[Kopparapu \& Barnes(2010)]{Kopparapu2010}
Kopparapu, R. \& Barnes, R. 2010. \apj, 716, 1336

\bibitem[Lebzelter et al.(2012)]{Lebzelter2012}
Lebzelter, T., Seifahrt, A., Uttenthaler, S. et al. 2012. \aap, 539, 25

\bibitem[Lepine \& Gaidos(2011)]{LG2011}
Lepine, S., \& Gaidos, E. 2011. \aj, 142, 15


\bibitem[Luinine et al.(2008)]{exoptf2008}
Lunine, J. I., Fischer, D., Hammel, H. B. 2008. {\it Astrobiology}, 8, Issue 5

\bibitem[Mahadevan et al.(2012)]{Suvrath2012}
Mahadevan, S., Ramsey, L., Bender, C. et al. 2012.{\it To appear in the proceedings of the SPIE2012 Astronomical Instrumentation and Telescopes conference}  arXiv:1209.1686

\bibitem[Manabe \& Wetherald(1967)]{MW1967}
Manabe, S., \& Wetherald, R. T. 1967. {\it Journal of Atmospheric Science}, 24, 241

\bibitem[Marshall \& Smith(1990)]{MS1990}
Marshall, B. R., \& Smith, R. C. 1990, Appl. Opt., 29, 71

\bibitem[Meadows \& Crisp(1996)]{MC1996}
Meadows, V. S., \& Crisp, D. 1996. {\it Journal of Geophysical Research}, 101, 4595

\bibitem[Mischna et al.(2000)]{Mischna2000}
Mischna, M.A., Kasting, J.F., Pavlov, A., Freedman, R., 2000. {\it Icarus}, 
145, 546

\bibitem[Mlawer et al.(1997)]{Mlawer1997}
Mlawer, E. J., Taubman, S. J., Brown, P. D. et al. 1997. {\it J. Geophys. Res.}, 102, 16663 

\bibitem[Nutzman \& Charbonneau(2008)]{Nutzman2008}
Nutzman, P., \& Charbonneau, D. 2008. {\it Publications of the Astronomical Society 
of the Pacific}, 120, 317

\bibitem[Orosz et al.(2012)]{Orosz2012}
Orosz, J. A., Welsh, W. F., Carter, J. A. et al. 2012. {\it Science}, 337, 1511

\bibitem[Paynter \& Ramaswamy (2011)]{PR2011}
Paynter, D. J. \& Ramaswamy, V. 2011. {\it Journal of
Geophysical Research-Atmospheres}, 116, D20302

\bibitem[Peale(1977)]{Peale1977}
Peale, S. J. 1977. Rotational histories of the natural satellites. In {\it Planetary Satellites}
, ed. J. A. Burns. Tucson, AZ: University of Arizona Press


\bibitem[Pepe et al.(2011a)]{Pepe2011a}
Pepe, F., Lovis, C., Segransan, D. et al. 2011a. \aap, 534, A58

\bibitem[Pepe et al.(2011b)]{Pepe2011b}
Pepe, F., Mayor, M., \& Lovis, C. et al. 2011b. in: A. Sozzetti,
M.G. Lattanzi, A.P. Boss (eds.), The astrophysics of planetary
systems: formation, structure, and dynamical evolution, IAU
Symp. 276, p. 13


\bibitem[Perrin \& Hartmann(1989)]{PH1989}
Perrin, M. Y., \& Hartmann, J. M. 1989, {\it JQSRT}, 42,311

\bibitem[Perryman et al.(2001)]{Perryman2001}
Perryman, M. A. C., de Boer, K.~S., Gilmore, G., et al. 2001, \aap, 369, 339

\bibitem[Pierrehumbert(2010)]{RayP2010}
Pierrehumbert, R. T. 2010. Principles of Planetary Climate, Cambridge University Press

\bibitem[Pierrehumbert(2011)]{RayP2011}
Pierrehumbert, R. T. 2011. \apjl, 726, L8

\bibitem[Pierrehumbert \& Gaidos(2011)]{PG2011}
Pierrehumbert, R. T. 2011. \apjl, 734, L13

\bibitem[Pollack et al.(1987)]{Pollack1987}
Pollack, J. B., Kasting, J.F., Richardson, S.M., Poliakoff, K., 1987. {\it Icarus}, 
71, 203


\bibitem[Ramirez et al.(2012a)]{Ramirez2012a}
Ramirez, R.M., Zugger, M.E., \& Kasting, J.F.,. 2012a. submitted {Nature Geoscience}


\bibitem[Ramirez et al.(2012b)]{Ramirez2012b}
Ramirez,  R., Kopparapu, R., \& Kasting, J. F. 2012b, submitted to {\it Icarus}

\bibitem[Robinson et al.(2011)]{Robinson2011}
Robinson, T., Meadows, V., Crisp, D., et al. 2011, {\it Astrobiology}, 11, 393

\bibitem[Rothman et al.(2009)]{Rothman2009}
Rothman, L. S., Gordon, I. E., Barber, A., et al. 2008. {\it JQSRT}, 110, 533

\bibitem[Rothman et al.(2010)]{Rothman2010}
Rothman, L. S., Gordon, I. E., Barber, A., et al. 2010. {\it JQSRT}, 111, 2139



\bibitem[Segura et al.(2002)]{Segura2002}
Segura, T. L., Toon, O. B., Colaprete, A., \& Zahnle, K. 2002. {\it Science}, 298, 1977

\bibitem[Segura et al.(2008)]{Segura2008}
Segura, T. L., Toon, O. B., \& Colaprete, A.  Journal of Geophysical Research, 113, E11007

\bibitem[Selsis et al.(2007a)]{Selsis2007a}
Selsis, F. et al. 2007a. {\it Icarus}, 191, 453

\bibitem[Selsis et al.(2007b)]{Selsis2007b}
Selsis, F. et al. 2007b. \aap, 476, 137

\bibitem[Shine et al.(2012)]{Shine2012}
Shine, K. P., Ptashnik, I. V., Radel, G. 2012. {\it Surveys in Geophysics}, 33, 535

\bibitem[Solomon \& Head(1991)]{SH1991}
Solomon, S. C., \& Head, J. W. 1991, {\it Science}, 252, 252

\bibitem[Spiegel et al.(2008)]{Spiegel2008}
Spiegel, D. S., Menou, K., \& Scharf, C. A. 2008. \apj, 681, 1609

\bibitem[Spiegel et al.(2009)]{Spiegel2009}
Spiegel, D. S., Menou, K., \& Scharf, C. A. 2009. \apj, 691, 596


\bibitem[Sykes(1952)]{Sykes1952}
Sykes, J. B. 1952. MNRAS, 3, 377

\bibitem[Terrien et al.(2012)]{Terrien2012}
Terrien, R. C. Mahadevan, S. Bender, C. et al. 2012. \apjl, 747, L38

\bibitem[Thomas \& Stamnes(2002)]{TS2002}
Thomas, G. E., \& Stamnes, K. 2002. Radiative Transfer in the Atmosphere and Ocean. Cambridge University Press

\bibitem[Tian et al.(2010)]{Tian2010}
Tian, F., Claire, M. W., Haqq-Misra, J. D., et al. 2010. Earth and Planetary Science Letters, 295, 412

\bibitem[Toon et al.(1989)]{Toon1989}
Toon, O., B., McKay, C. P., Ackerman, T., P., \& Santhanam, K. 1989, {\it J. Geophys. Res.}, 94,16287

\bibitem[Traub(2012)]{Traub2012}
Traub, W. A. 2012. \apj, 745, article id. 20

\bibitem[Tuomi et al.(2012a)]{Tuomi2012a}
Tuomi, M., Anglada-Escude, G., Gerlach, E. et al. 2012a. \aap. 549. id.A48

\bibitem[Tuomi et al.(2012b)]{Tuomi2012b}
Tuomi, M., Jones, H. R. A., Jenkins, J. et al. 2012b. \aap, accepted arXiv:1212.4277

\bibitem[Udry et al.(2007)]{Udry2007}
Udry, S., Bonfils, X., Delfosse, X. et al. 2007. \aap. 469, L43

\bibitem[Underwood et al.(2003)]{Underwood2003}
Underwood D. R., Jones, B. W., \& Sleep, P. N. 2003. International Journal of Astrobiology,
2, 289

\bibitem[Vardavas \& Carver(1984)]{VC1984}
Vardavas, I. M., \& Carver. J. H. {\it Planet. Space Sci.}, 32, 1307

\bibitem[Vogt et al.(2010)]{Vogt2010}
Vogt, S. S., Butler, R. P., \& Rivera, E. J. et al. 2010. \apj, 723, 954

\bibitem[Vogt et al.(2012)]{Vogt2012}
Vogt, S. S.,  Butler, P., \& Haghighipour, N. 2012. {\it Astronomische Nachrichten}, 333, 561

\bibitem[Von Paris et al.(2010)]{vparis2010}
Von Paris, P. et al. 2010. \aap, 522, A23

\bibitem[Von Paris et al.(2011a)]{vparis2011a}
Von Paris, P., Gebauer, S., Godolt, M., Rauer, H., \& Stracke, B.  2011a. \aap, 532, id.A58


\bibitem[Welsh et al.(2012)]{Welsh2012}
Welsh, W., Orosz, J. A., Carter, J. A. et al. 2012. {\it Nature}, 481, 475

\bibitem[Whitmire et al.(1998)]{Whitmire1998}
Whitmire, D. P., Matese, J. J., Criswell, L. 1998. {\it Icarus}, 132, 196

\bibitem[Williams \& Pollard(2002)]{WP2002}
Williams, D. M., \& Pollard, D. 2002. {\it Int. J. Astrobiol.}, 1, 61

\bibitem[Wordsworth et al.(2010)]{Wordsworth2010}
Wordsworth, R., Forget, F., \& Eyment, V. 2010, {\it Icarus}, 210, 2, 992

\bibitem[Wordsworth et al.(2011)]{Wordsworth2011}
Wordsworth, R. D., Forget, F., Selsis, F. et al. 2011. \apj, 733, L48


\bibitem[Wright et al.(2011)]{Wright2011}
Wright, J. T., Fakhouri, O., Marcy, G. W. et al. 2011. {\it PASP}, 123, 412

\bibitem[Zsom et al.(2012)]{Zsom2012}
Zsom, A., Kaltenegger, L., \& Goldblatt, C. 2012. To appear in Icarus, arXiv:1208.5028

\end{thebibliography}
\end{document}